\documentclass[epj-spec]{svjour}
\def\makeheadbox{}
\usepackage{url}
\usepackage{amstext}
\usepackage{bm}
\usepackage{graphicx}
\newcommand{\onlinecite}[1]{\cite{#1}}
\begin{document}
  \title{Dynamical mean-field approach to materials with strong electronic correlations}
  \author{%
    J.~Kune\v{s},\inst{1}
    I.~Leonov,\inst{2}
    M.~Kollar,\inst{2}
    K.~Byczuk,\inst{3}
    V.~I. Anisimov,\inst{4}
    and D.~Vollhardt\inst{2,}\thanks{\email{dieter.vollhardt@physik.uni-augsburg.de}}} 
  \institute{%
    Institute of Physics, Academy of Sciences of the Czech Republic, Cukrovarnicka 10,
    Praha~6 16253, Czech Republic
    \and
    Theoretical Physics III, Center for Electronic Correlations and
    Magnetism, Institute of Physics, University of Augsburg, 86135
    Augsburg, Germany
    \and
    Institute of Theoretical Physics, University of Warsaw,
    ul.\ Ho\.za~69, 00-681 Warszawa, Poland
    \and
    Institute of Metal Physics, Russian Academy of Sciences, 620041
    Yekaterinburg GSP-170, Russia
  }

  \abstract{We review recent results on the properties of materials
    with correlated electrons obtained within the LDA+DMFT approach, a
    combination of a conventional band structure approach based on the
    local density approximation (LDA) and the dynamical mean-field
    theory (DMFT). The application to four outstanding problems in
    this field is discussed: (i) we compute the full valence band
    structure of the charge-transfer insulator NiO by explicitly
    including the $p$-$d$ hybridization, (ii) we explain the origin for
    the simultaneously occuring metal-insulator transition and
    collapse of the magnetic moment in MnO and Fe$_2$O$_3$, (iii) we
    describe a novel GGA+DMFT scheme in terms of plane-wave
    pseudopotentials which allows us to compute the orbital order and
    cooperative Jahn-Teller distortion in KCuF$_3$ and LaMnO$_3$, and
    (iv) we provide a general explanation for the appearance of kinks
    in the effective dispersion of correlated electrons in systems
    with a pronounced three-peak spectral function without having to
    resort to the coupling of electrons to bosonic excitations.  These
    results provide a considerable progress in the fully microscopic
    investigations of correlated electron materials.
  }

  \maketitle

  \markboth{}{}

  \section{Introduction}

  The theoretical understanding of materials with strong effective
  interactions between the electrons is one of the most challenging
  areas of current research in condensed matter physics \cite{Rev1}.
  Strong electronic interactions are encountered in materials with
  open $d$ and $f$ shells, such as the transition metals Ti, V, Fe and
  their oxides or rare--earth metals such as Ce, where electrons
  occupy narrow orbitals.  This spatial confinement enhances the
  effect of the Coulomb interaction between the electrons, making them
  ``strongly correlated''. The interplay between the spin, charge and
  orbital degrees of freedom of the correlated $d$ and $f$ electrons
  and the lattice degrees of freedom leads to a multitude of unusual
  ordering phenomena at low temperatures. Consequently, strongly
  correlated electron systems are often exceedingly sensitive to small
  changes in the temperature, pressure, magnetic field, doping, and
  other control parameters. This results, for example, in large
  changes of the resistivity across metal-insulator transitions or
  upon the application of a magnetic field of the volume across phase
  transitions, and of the effective electronic masses; electronic
  correlations are also essential for an understanding of high-temperature
  superconductivity.  Correlated electron materials often
  reveal rich phase diagrams originating from the interplay between
  electronic and lattice degrees of freedom \cite{Rev1,Rev2,Rev3}.
  These compounds are particularly interesting in view of possible
  technological applications. Namely, their properties can be employed
  to construct materials with new functionalities for future
  technological applications.

  The electronic properties of many materials can be computed using
  density functional theory methods, e.g., in the local density
  approximation (LDA) \cite{LDA}, the generalized gradient
  approximation (GGA) \cite{PB96,TB09}, or using the so-called LDA+U
  method \cite{AZ91,LA95,AA97}.  Applications of these approaches
  often describe the physical properties of simple elements and
  intermetallic compounds, and of some insulators, quite accurately.
  Moreover, they permit to make correct qualitative predictions of the
  magnetic, orbital, and crystal structures of solids where the
  equilibrium (thermodynamic) structures are determined by
  simultaneous optimization of the electron and lattice systems
  \cite{RevPWSCF1,RevPWSCF2,RevWIEN2k,RevVASP}.
  However, these methods usually fail to describe the correct
  electronic and structural properties of electronically correlated
  \emph{paramagnetic} materials.
  Hence the computation of electronic, magnetic, and structural
  properties of strongly correlated paramagnetic materials remains a
  great challenge.

  Here the recent combination of conventional band structure theory
  \cite{LDA} and dynamical mean-field theory
  \cite{DMFT1,DMFT2,DMFT3,DMFT4,DMFT5}, the so-called LDA+DMFT
  computational scheme
  \cite{DMFTmeth1,DMFTmeth2,DMFTmeth3,DMFTmeth4,DMFTmeth5,DMFTmeth6}
  has led to a powerful new tool for the investigation of strongly
  correlated compounds both in their paramagnetic and magnetically
  ordered states.  In particular, it has already provided important
  insights into the spectral and magnetic properties of correlated
  electron materials
  \cite{DMFTcalc1,DMFTcalc2,DMFTcalc3,DMFTcalc4,DMFTcalc5,HM01,MH03,AB06,SKA01,DSK03,SK04,DMFTcalc+1,DMFTcalc+2,DMFTcalc+3,DMFTcalc+4,DMFTcalc+5,kun07a,kun07b,KL08,fe2o3}
  especially in the vicinity of a Mott
  metal-insulator transition as encountered in transition metal oxides.

  In this paper we review several topics related to the physics of
  correlated electron materials where significant progress was made
  during the last three to four years. In particular we describe the
  application of the LDA+DMFT approach to charge-transfer materials
  where --- in contrast to the early transition metal oxides ---
  additional complexities from the presence of $p$ bands arise. We
  apply the scheme to study the single-particle spectrum of the
  prototypical charge-transfer insulator NiO. By explicitly including
  the O-$p$ orbitals and their hybridization with Ni-$d$ orbitals we
  obtain a unified description of the full spectrum (Sec.~\ref{sec:2}). A
  second topic is the investigation of the influence of pressure on
  correlated electronic materials and of the collapse of the magnetic
  moment. Indeed, in numerous transition-metal oxides the
  pressure-induced metal-insulator transition is accompanied by a
  collapse of the magnetic moments, i.e., a transition from a high
  spin state to a state with low spin. Such a transition is observed
  for example in MnO and Fe$_2$O$_3$. Our results obtained with
  LDA+DMFT for MnO are described in Sec.~\ref{sec:3}. A breakthrough in the
  application of the LDA+DMFT scheme concerning the computation of
  correlation induced atomic displacements and structural
  transformations is presented in Sec.~\ref{sec:4}.  Namely, by formulating a
  LDA/GGA+DMFT scheme implemented with plane-wave pseudopotentials it
  has been possible to describe and understand the structural
  transformation of paramagnetic solids due to electronic correlation
  effects. Finally, we present a general explanation for the
  appearance of kinks, i.e., sudden changes in the slope of the
  effective electronic dispersion of the correlated electrons. By a
  detailed analytic investigation of the Green function and the
  self-energy for correlated systems with a pronounced three-peak
  spectral function we are able to explain the occurrence of kinks
  without having to resort to any explicit coupling of electrons to
  bosonic excitations (Sec.~\ref{sec:5}).

  \section{Charge-transfer compounds}
  \label{sec:2}

  In the so-called charge-transfer (CT) materials
  the ligand orbitals play an important role since they 
  determine the low-energy physics of these systems.  The concept of a
  CT insulator was introduced in the mid 1980's by Zaanen, Sawatzky
  and Allen (ZSA) in their classification of transition-metal oxides
  (TMOs) and related compounds~\cite{ZSA85}.  In the early TMOs the
  ligand $p$ band is located well below the transition metal $d$ band
  and thus plays a minor role in the low energy dynamics. Such a case,
  called Mott-Hubbard system in the ZSA scheme, is well described by a
  multi-band Hubbard model.  On the other hand, the late TMOs are of
  the CT type where the $p$ band is located between the interaction
  split $d$ band. A more general Hamiltonian where the $p$ states are
  explicitly included is then needed, which can be viewed as a
  combination of multi-band Hubbard and Anderson lattice models.  The
  results of calculations using the static mean-field approximation
  such as LDA+U~\cite{AZ91,LA95,AA97} suggested that the treatment of
  the dynamical correlations and the ligand--transition-metal
  hybridization might be important to properly describe the
  excitations in the CT materials.

  NiO is a well-studied prototype of a CT insulator. It is a type II
  anti-ferromagnet ($T_N$ $=$ 523~K) with a magnetic moment of almost
  2$\mu_B$ and a large gap surviving well above $T_N$.  Its
  photoemission spectrum~\cite{saw84,eastman,she90,she91}, which is
  our main concern, has several characteristic features: (i) a broad
  high-energy peak with little dispersion dominated by Ni $d$
  character, (ii) a flat low-energy band with substantial amount of Ni
  $d$ character, (iii) dispersive O $p$ bands at intermediate
  energies, and (iv) a charge gap of 3-4~eV. All of these feature are
  essentially the same in both the antiferro- and the paramagnetic
  phase~\cite{tje96}.  Although further details can be resolved in the
  spectra of NiO we will focus on these four features here.

  The LDA+DMFT calculations proceeds in two independent steps: the
  construction of the effective Hamiltonian from converged LDA
  calculation, and the subsequent solution of the corresponding DMFT
  equations.  We use the Wannier function basis~\cite{MV97,AK05} for
  an eight-band $p$-$d$ Hamiltonian
  \begin{eqnarray}
    H&=&
    \sum_{\bm{k},\sigma}\bigl(h_{\bm{k},\alpha\beta}^{dd}
    d_{\bm{k}\alpha\sigma}^{\dagger}
    d_{\bm{k}\beta\sigma}+h_{\bm{k},\gamma\delta}^{pp}p_{\bm{k}
      \gamma\sigma}^{\dagger}
    p_{\bm{k}\delta\sigma}+
    h_{\bm{k},\alpha\gamma}^{dp}
    d_{\bm{k}\alpha\sigma}^{\dagger}
    p_{\bm{k}\gamma\sigma}+h_{\bm{k},\gamma\alpha}^{pd}
    p_{\bm{k}\gamma\sigma}^{\dagger}
    d_{\bm{k}\alpha\sigma}\bigr)
    \nonumber\\&&
    +\sum_{i,\sigma,\sigma'}
    U_{\alpha\beta}^{\sigma\sigma'}n^d_{i\alpha\sigma}n^d_{i\beta\sigma'}.\label{eq:ham}
  \end{eqnarray}
  Here $d_{\bm{k}\alpha\sigma}$ and $p_{\bm{k}\gamma\sigma}$ are
  Fourier transforms of $d_{i\alpha\sigma}$ and $p_{i\gamma\sigma}$,
  which annihilate the $d$ or $p$ electron with orbital and spin
  indices $\alpha\sigma$ or $\gamma\sigma$ in the $i$th unit cell, and
  $n^d_{i\alpha\sigma}$ is the corresponding occupation number
  operator.  The elements of the $U_{\alpha\beta}^{\sigma\sigma'}$ matrix
  are parameterized by $U$ and $J$. The constrained LDA calculation
  yields $U$ $=$ 8~eV and $J$ $=$ 1~eV~\cite{AZ91}.  To account for the
  Coulomb energy already present in LDA we subtract a double counting
  correction from the $dd$-diagonal elements of the LDA Hamiltonian:
  \begin{equation}
    h^{dd}_{\bm{k},\alpha\beta}=\tilde{h}^{dd}_{\bm{k},\alpha\beta}
    (\bm{k})-(N_{\text{orb}}-1)
    \bar{U}n_{\text{LDA}}\delta_{\alpha\beta}
  \end{equation}
  where $n_{\text{LDA}}$ is the average LDA occupation per orbital and
  $N_{\text{orb}}$ $=$ 10 is the total number of orbitals within the
  shell.  With the kinetic and interaction parts of the Hamiltonian
  specified we solve the DMFT equations on the imaginary frequency axis, a
  key part of which is the auxiliary impurity problem treated with
  the quantum Monte-Carlo (QMC) method~\cite{HF86}.  The results reported
  here were calculated at $T$ $=$ 1160~K.  To obtain the single-particle
  spectral functions an analytic continuation to real frequencies is
  performed using the maximum entropy method~\cite{mem}.

\begin{figure}
    \includegraphics[angle=270,width=\columnwidth,clip]{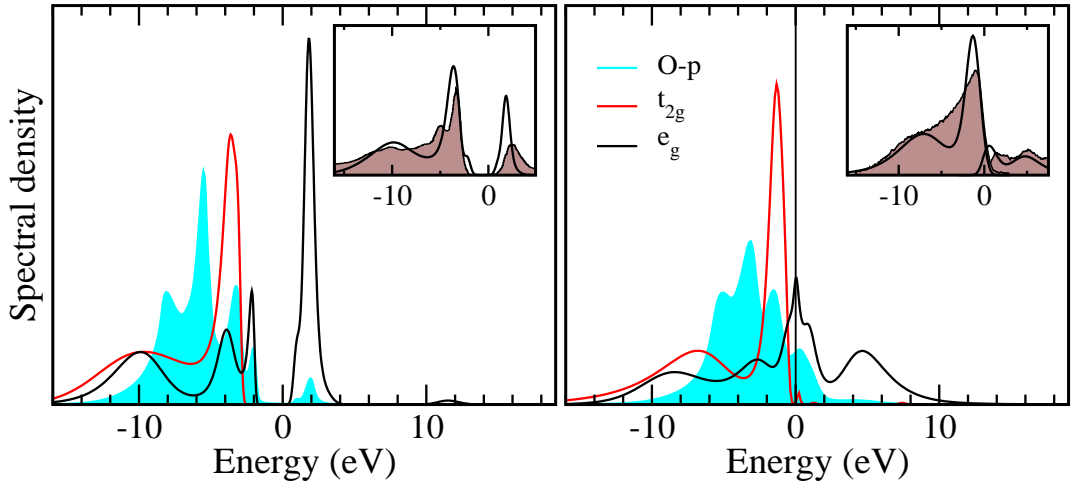}
    \caption{\label{fig:nio-pes}(color online) Left panel: the
      orbitally resolved single-particle spectrum of NiO with
      contributions from Ni $d$-$e_g$ (black), Ni $d$-$t_{2g}$ (red),
      and O $p$ (blue shaded) states. The inset shows comparison of
      the total Ni $d$ spectral density (with Gaussian broadening of
      0.6~eV) (black) to the XPS spectra (brown shaded) of
      Ref.~\onlinecite{saw84}. Right panel: the orbitally resolved
      single-particle spectrum of hole doped NiO ($n_h$ $=$ 0.6). The
      total Ni $d$ spectrum (black) is compared to the experimental
      photoemission and inverse-photoemission spectra of
      Li$_{0.4}$Ni$_{0.6}$O~\cite{linio} (brown shaded).}
  \end{figure}

  We start the discussion of our numerical results with the
  single-particle spectra of stoichiometric NiO. Using the full
  $p$-$d$ Hamiltonian we are able to cover the entire valence and
  conduction bands spectra. In the left panel of
  Fig.~\ref{fig:nio-pes} we show the local ($\bm{k}$-integrated)
  spectral densities corresponding to Ni $d$ $e_g$ and $t_{2g}$, as
  well as O $p$ electrons. In the inset we compare the total Ni $d$
  spectral density to the experimental data of Sawatzky and
  Allen~\cite{saw84}.  The theoretical spectrum very well reproduces
  the experimental one, including the size of the gap, the $d$
  character of the conduction band, the broad $d$ peak at $-9$~eV, the
  position of the $p$ band, and the strong $d$ contribution at the top
  of the valence band. The dominant feature of the valence spectrum is
  a distribution of spectral weight between the broad peak at high energy
  and sharp peaks at low energy.  The spectral weight
  distribution is a combined effect of the $p$-$d$ hybridization and
  the electronic correlation in the Ni $d$ shell as was explained by
  Fujimori {\it et al.}~\cite{fuj84} in terms of eigenstates of a
  NiO$_6$ cluster, and Sawatzky and Allen~\cite{saw84}.  Emitting a
  $d$ electron from the $d^8$ state the system can reach either the
  $d^7$ final state (high-energy peak) or a $d^8\underline{L}$ final
  state (sharp low-energy peak), with a ligand hole, due to the
  $p$-$d$ electron transfer.

  Virtually all electronic structure computational methods have been
  applied to NiO, often claiming success in describing its
  photoemission spectrum.  However, these claims are often based 
  on picking only a particular aspect of the problem. Putting aside the
  fact that static methods such as LDA+U~\cite{AZ91,LA95,AA97} and
  generalized density functionals~\cite{PB96,TB09} as well as
  $GW$~\cite{FSK04,LR05} provide gapped spectra only in the AFM phase, the
  main difficulty appears to be capturing the distribution of spectral
  weight between the high-energy $d^7$ peak of Ni $d$ character and
  the low-energy $d^8\underline{L}$ band with mixed Ni $d$ - O $p$
  composition. For example, the high-energy Ni $d$ peak is completely
  missing in the calculations with the recently introduced
  density~\cite{PB96,TB09} and density matrix~\cite{shar09}
  functionals and also in $GW$
  calculations~\cite{FSK04,LR05,kob08,rod09} for reasons explained in
  Ref.~\onlinecite{ary95}. Also LDA+U calculations with small $U$ miss
  the high-energy peak~\cite{coco05}. On the other hand, for larger
  $U$ values the $d$ character is missing in the low-energy
  peak~\cite{AZ91,rod09}. The early LDA+DMFT calculations of Ren {\it
    et al.}~\cite{ren06} also do not yield qualitatively correct
  valence band since the O $p$ states were not included in the
  effective Hamiltonian.  On the other hand many-body techniques such
  as cluster exact diagonalization~\cite{fuj84}, three-body scattering
  technique~\cite{man94} or
  DMFT~\cite{kun07a,kun07b,eder07,yin08,miu08}, which are capable of
  generating new poles in the single-particle self-energy, describe
  the distribution of the valence spectral weight in 
  agreement with experiment.

\begin{figure}
    \includegraphics[width=\columnwidth,clip]{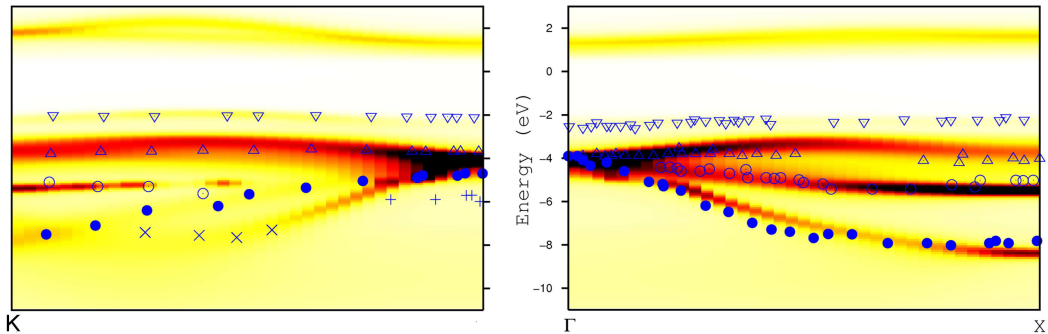}
    \caption{\label{fig:arpes} The ${\bm{k}}$-resolved total spectral
      function $A({\bm{k}},\omega)$ along the $K-\Gamma$ (left panel)
      and $\Gamma-X$ (right panel) lines in the Brillouin zone
      depicted as a contour plot. The symbols represent the
      experimental bands of Shen {\it et al.}~\cite{she91}. The
      theoretical gap edge was aligned with the experimental one.  }
  \end{figure}

  To further investigate the electronic structure of NiO we have
  calculated the $\bm{k}$-resolved spectra, which determine the correlated bandstructure.  In
  Fig.~\ref{fig:arpes} we compare the theoretical bands, represented
  by the ${\bm{k}}$-dependent spectral density $A({\bm{k}},\omega)$,
  along the $X-\Gamma$ and $\Gamma-K$ lines in the Brillouin zone with
  ARPES data of Refs.~\onlinecite{she90,she91}.  Both theory and
  experiment exhibit two relatively flat bands at $-2$ and $-4$~eV
  followed by several dispersive bands in the $-4$ to $-8$~eV range and a
  broad incoherent peak around $-10$~eV. Overall we find a good
  agreement.  The deviations around the $\Gamma$ point in the right
  panel of Fig.~\ref{fig:arpes} are due to the inaccuracy in the
  location of the $\Gamma$ point in the off-normal-emission
  experiment~\cite{she91}.  The crosses near the $\Gamma$ point mark a
  weak band which was interpreted as a consequence of AFM
  order~\cite{she91} and is therefore not expected to be found in the
  paramagnetic phase investigated here.  Additional weak structures in
  the uppermost valence band (not shown here) were identified in the
  spectra of Ref.~\onlinecite{she91}. To assess their importance the
  reader is referred to the original experimental spectra.

  The effect of doping holes to a CT insulator is of prime importance,
  particularly because the high-temperature superconductivity is
  widely believed to arise from this effect. Besides that, the $p$
  character of the doped holes is a defining feature of CT materials.
  In NiO hole doping can be achieved by Li substitution.
  Li$_x$Ni$_{1-x}$O was studied in the doping range
  $x$ $=$ 0.02-0.4~\cite{linio}.  Starting from the undoped system the
  replacement of $x$ Ni$^{2+}$ ions by Li$^{1+}$ ions introduces on
  average $n_h=x/(1-x)$ holes per Ni site.  In the following, we
  neglect the substitutional disorder and study the Hamiltonian of NiO
  at various non-stoichiometric fillings.  As crude as this
  approximation may be we believe that the essential physics of
  $p$-$d$ spectral weight transfer is captured correctly.
  \begin{table}
    \centering

    \caption{\label{tab:1} Orbital occupancies and the local moment on the Ni site for
      different hole dopings.}
    \begin{tabular}{c|ccccc}
      $n_h$ & $n_{e_g}$ & $n_{t_{2g}}$ & $n_p$ & $M_s$ & $M_{\text{eff}}$ \\
      \hline
      0 & 0.547 & 1.000 & 0.969 & 1.85 & 1.82\\
      0.6 & 0.531 & 0.994 & 0.885 & 1.61 & 1.27\\
      1.2 & 0.530 & 0.980 & 0.800 & 1.45 & 0.89\\
    \end{tabular}
  \end{table}
  In Table \ref{tab:1} we compare the orbital occupancies and
  local spin moment for different hole dopings.
  While a substantial $d$ spectral density is observed at the top of NiO valence
  band the doped holes reside almost exclusively in oxygen $p$ orbitals.
  This apparent contradiction to the experience from non-interacting electron
  systems reflects
  the importance of many-body effects.

  Yet another demonstration of electronic correlations is a substantial
  reconstruction of the spectral function, in particular the $e_g$ one,
  shown in the left panel of Fig.~\ref{fig:nio-pes}. Most notably
  the Mott gap is filled, while the Hubbard subbands remain as distinct features
  with the spectral weight reduced as compared to pure NiO.
  This is also observed in experiment, as shown in the inset.
  To understand the spectrum it is useful to think about the doping in two steps.
  First, the holes are doped into a system with no $p$-$d$ hybridization.
  For the present heavy doping this simply amounts to shifting the chemical
  potential deep into the non-interacting $p$ band. Second, the $p$-$d$ hybridization
  is turned on and the system is essentially described by the periodic Anderson model
  with a well known spectrum similar to that for the $e_g$ orbital \cite{JWA}.

\begin{figure}
    \centering
    \includegraphics[height=0.75\columnwidth,angle=270,clip]{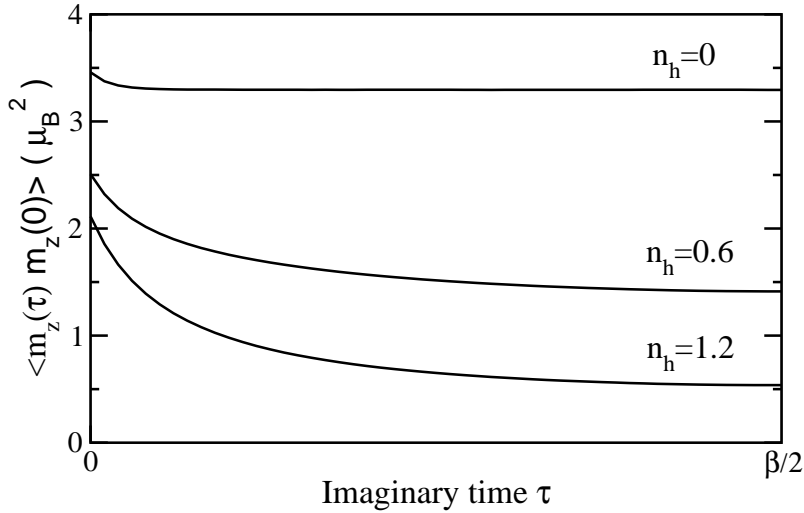}
    \caption{\label{fig:mtaum}
      The imaginary time spin-spin correlations $\langle m_z(\tau)m_z(0)\rangle$
      for various hole dopings. The calculations were performed at 1160~K ($\beta$ $=$ 10~eV$^{-1}$).}
  \end{figure}

  While the $d$ occupancy is barely affected by the hole doping, the
  behavior of the local spin moment changes. The instantaneous moment
  $M_s=\sqrt{\langle \hat{m}_z^2 \rangle}$ is reduced due to the suppression
  of the dominant $d^8$, $m_z=2$ contributions in favor of $d^6$,
  $d^7$, and $d^8$ contributions with $m_z=0$ and $m_z=1$.  Even more
  pronounced is the effect of increased screening of the local moment,
  reflected in faster decay of imaginary time spin-spin correlations
  (see Fig.~\ref{fig:mtaum}), leading to a significant departure of
  the screened moment $M_{\text{eff}}=\sqrt{T\chi_{\text{loc}}}$ from its
  instantaneous value.

  To summarize, we have presented LDA+DMFT calculations on NiO and its
  hole doped relatives using a Hamiltonian spanning Ni $3d$ and O $2p$
  bands. It appears that the main limitation of the truly first
  principles character of LDA+DMFT studies is the need for a
  double-counting correction, the universal form of which is not known,
  and perhaps does not exist. We also mention the density-density form
  of the electron-electron interaction used here, a rather technical
  approximation, which can be avoided with the most recent CT-QMC
  implementations~\cite{laue09} or exact diagonalization~\cite{ishida10}. Despite these approximations we were
  able to obtain the main characteristics of the photoemission and
  inverse-photoemission spectra of both pure and hole-doped NiO. The
  computational methods claiming success for NiO and CT compounds in
  general should be able to capture substantial $d$ spectral weight at
  the top of the valence band together with the incoherent lower
  Hubbard band, and, perhaps most importantly, should be able to
  describe the peculiar behavior upon hole doping.

  \section{Metal-insulator and spin transitions}
  \label{sec:3}

  A very interesting type of metal-insulator transition (MIT) is the
  pressure driven transition accompanied by a change of the local spin
  state (high spin (HS) to low spin (LS) transition) seen in
  MnO~\cite{yoo-05}, BiFeO$_3$~\cite{GSL08} or Fe$_2$O$_3$.
  An understanding of the pressure-driven HS-LS transition and its
  relationship to the MIT and to structural and/or volume changes is relevant
  to a broader class of oxides~\cite{lyu09}, often with geophysical
  implications~\cite{cohen97}.

  We have studied two isoelectronic materials which exhibit the above
  behavior under pressure, hematite ($\alpha$-Fe$_2$O$_3$)~\cite{fe2o3} and
  manganosite (MnO)~\cite{KL08}.  At ambient conditions, Fe$_2$O$_3$ is an
  antiferromagnetic (AFM) insulator ($T_N$ $=$ 956~K) with the
  corundum structure~\cite {Sgull-51} while MnO is antiferromagnetic
  only below 118~K and possesses rock-salt structure. In both
  compounds the transition metal (TM) ions have formal valence state
  $d^5$ and are octahedrally coordinated with oxygen, which gives rise
  to the characteristic $e_g$-$t_{2g}$ splitting.  Photoemission
  spectroscopy (PES) at zero pressure~\cite{Fujimori-86,KK02,lad-89}
  classified Fe$_2$O$_3$ as a charge-transfer insulator with the
  charge gap of 2.0-2.7~eV inferred from the electrical conductivity
  data~\cite{mochizuki,KL85}.  Under pressure, a first-order phase
  transition is observed at approximately 50~GPa (82\% of the
  equilibrium volume) with the specific volume decreasing by almost
  10\% and the crystal symmetry being reduced (to the Rh$_2$O$_3$-II
  structure)~\cite{Pasternak-99,Rozenberg-02,Liu-03}.  The
  high-pressure phase is characterized by a metallic conductivity and
  the absence of both magnetic long-range order and the HS local
  moment~\cite{Pasternak-99}. Badro {\it et al.}  showed that the
  structural transition actually precedes the electronic transition,
  which is, nevertheless, accompanied by a sizable reduction of the
  bond lengths~\cite{Badro-02}.

  \begin{figure}
    \centering
    \includegraphics[height=0.85\columnwidth,angle=270,clip]{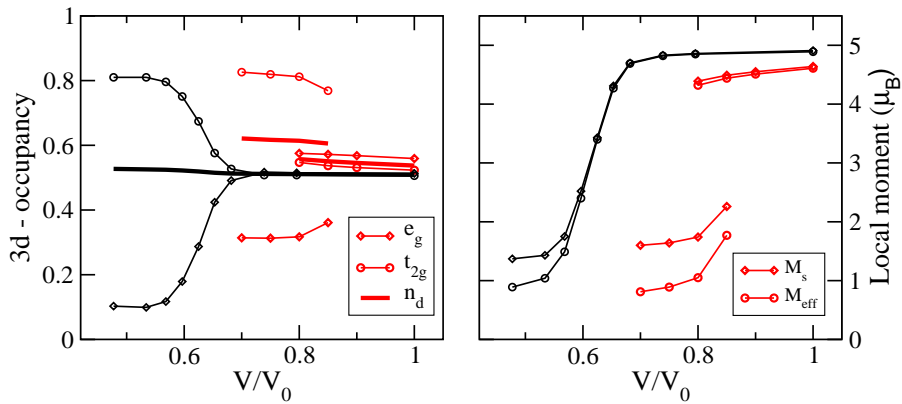}
    \caption{\label{fig:mom} Left panel: the orbital-resolved
      occupancies of the 3$d$ shell of Fe in Fe$_2$O$_3$ (red) and Mn
      in MnO (black). The thick lines show the average occupancy $n_d$ per
      orbital. Right panel: the local spin moments in Fe$_2$O$_3$
      (red) and MnO (black) measured as instantaneous moment
      (diamonds) and screened moment (circles). The symbols mark the
      actual numerical results, the lines are guides to the eye. 
      $V_0$ is the volume at ambient pressure. Adapted from \cite{KL08}.}
  \end{figure}

  MnO shows a similar behavior. The shock data~\cite{noguchi}, and
  then Raman and optical studies~\cite{mita,mita2}, had identified a
  transformation in MnO in the vicinity of 90-105 GPa.
  Transport~\cite{patterson}, magnetic~\cite{patterson}, structural
  and spectroscopic~\cite{yoo,rueff}, and reflectivity~\cite{mita}
  data all point to a first-order, insulator-metal Mott transition
  near 100 GPa with volume ($v=V/V_0$, $V_0$ $=$ volume at ambient pressure) collapse
  $v$=0.68$\rightarrow$0.63, and moment collapse (from $\sim$5$\mu_B$
  to 1$\mu_B$ or less~\cite{yoo,rueff}).  The structural data indicate
  a B1$\rightarrow$B8 change just before the Mott transition, which
  thus occurs within the B8 (NiAs) phase rather than the B1 (NaCl)
  phase.  Since the local environment of the Mn ion remains the same,
  this structural change is not expected to have much effect on the
  Mott transition.

  Our calculations proceed along the lines described in the preceding
  section: i) construction of the multi-band Hubbard Hamiltonian
  spanning the TM $d$ and ligand $p$ bands with local Coulomb
  interaction within the TM $d$ shell in the density-density (Ising)
  approximation, ii) iterative solution of the DMFT equations on the
  imaginary axis using QMC, iii) computation of
  observables of interest. This procedure has been repeated for
  several lattice parameters corresponding to pressures up to several
  100~GPa. In the case of MnO we employed the Hirsch-Fye QMC
  algorithm  available at the time, for which  the calculations
  were limited to the temperature $T$ $=$ 1160~K. For the Fe$_2$O$_3$
  study the continuous-time QMC algorithm (hybridization expansion) was
  used, which allowed us to reach room temperature and study the
  anti-ferromagnetic ordering. In both cases we had to introduce
  special Monte-Carlo moves to ensure ergodic simulation for
  parameters close to the spin transition.  The calculated observables
  include the single-particle spectral density to study the formation
  of the charge gap, local orbital occupancies $\langle n_i\rangle$
  and static local density-density correlations $\langle n_i n_j
  \rangle$ to investigate the orbital redistributions, instantaneous local
  moments and interaction energy, and local spin susceptibility to
  determine the screening of the local spin moments.

\begin{figure}
    \centering
    \includegraphics[height=0.75\columnwidth,angle=270,clip]{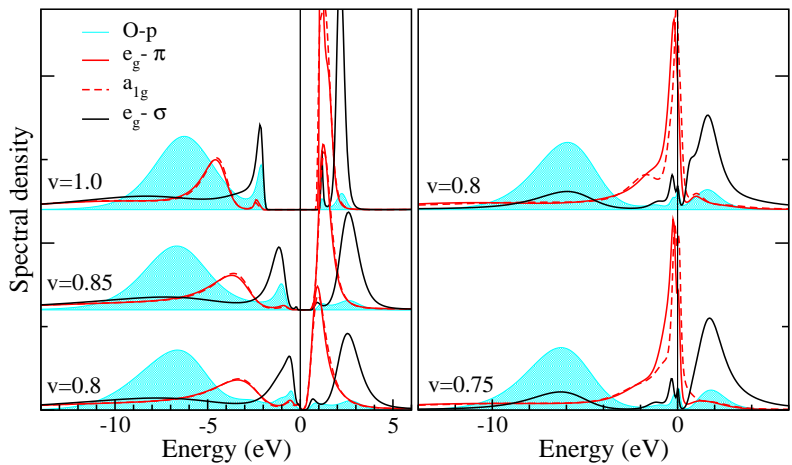}
    \caption{\label{fig:spec-fe2o3} The single-particle spectra of
      Fe$_2$O$_3$ at various specific volumes ($T$ $=$ 580~K) \cite{fe2o3}. The HS
      solutions are shown in the left and LS solutions in the right
      panel. The Fe $d$ spectra are resolved into the $e_g$ (called
      $e_g^{\sigma}$ in the legend) and $t_{2g}$ (further split in
      $a_{1g}$ and $e_g^{\pi}$ by distortion from precise octahedral
      symmetry) contributions. The O $p$ spectra are marked by blue
      shading.}
  \end{figure}

\begin{figure}
    \centering
    \includegraphics[height=0.75\columnwidth,angle=270,clip]{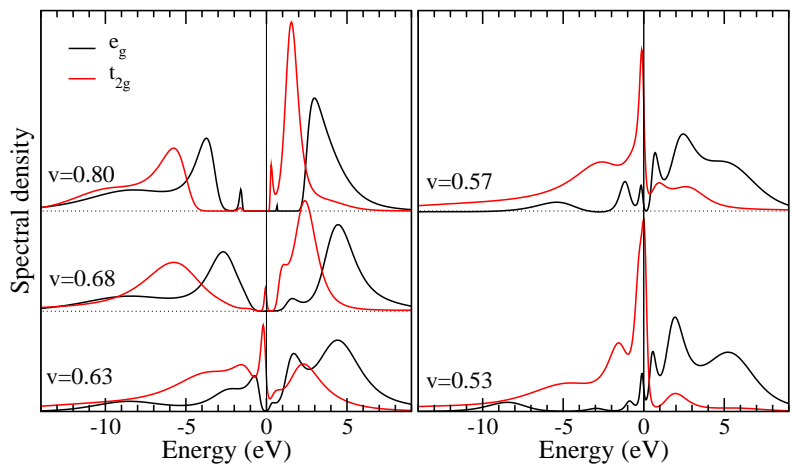}
    \caption{\label{fig:spec-mno} The single-particle spectra of MnO
      at various specific volumes ($T$ $=$ 1160~K) in the vicinity of
      the spin transition~\cite{KL08}. The Mn $d$ spectra are resolved into the
      $e_g$ and $t_{2g}$ contributions, O $p$ spectra are not shown.}
  \end{figure}

  In Fig.~\ref{fig:mom} we show the evolution of the local moments and
  the orbital occupancies as functions of specific volume.  We use two
  different quantities to characterize the local moment: (a) the mean
  instantaneous moment defined as an equal time correlation function
  $M_s=\sqrt{\langle \hat{m}_z^2 \rangle}$ and (b) the screened local
  moment defined through the local spin susceptibility
  $M_{\text{eff}}=\sqrt{T\chi_{\text{loc}}}$.  These two different definitions of the
  local moment yield almost the same $T$-independent values in materials
  with Curie-Weiss behavior. The two studied materials exhibit clear
  similarities. In particular, the HS state with $M_s
  \sim 5$ and orbital occupancies corresponding to the complete spin
  polarization of a half-filled $d$ shell dictated by the first Hund's
  rule, the LS state characterized by emptying of the $e_g$ orbitals,
  substantial reduction of the local moment and the screening of the
  local moment reflected in the difference between $M_s$ and
  $M_{\text{eff}}$. Furthermore it turns out that the charge gap disappears
  in the LS phase (see Figs. \ref{fig:spec-fe2o3} and
  \ref{fig:spec-mno}).

  Nevertheless, there are also several differences between MnO and
  Fe$_2$O$_3$. Their orbital occupancies differ quantitatively,
  especially the $e_g$ occupancies in the LS phase are substantially
  larger in Fe$_2$O$_3$. This is a result of more covalent character
  of this compound, i.e., stronger $p$-$e_g$ mixing. Another
  consequence of the strong covalent $p$-$e_g$ bonding is a transfer
  of charge between Fe and O upon the HS-LS transition, reflected in
  the difference of the total occupancies $n_d$ in the two phases.
  There is also a qualitative difference between the transitions in
  MnO and Fe$_2$O$_3$. In MnO we find a smooth crossover characterized
  by a unique solution for all specific volumes with intermediate
  magnetic moments, and only when the lattice is included we observe a
  phase separation into high-volume HS phase and low-volume LS phase
  in the crossover region.  In Fe$_2$O$_3$ we observe an
  electronically first-order transition with hysteresis.

  Next, we compare the evolution of the single-particle spectral
  densities (see Figs. \ref{fig:spec-fe2o3} and \ref{fig:spec-mno}) as
  the volume is reduced.  In both cases we observe the closing of the HS
  charge gap as the material is compressed, and LS spectra that bear
  clear resemblance to the non-interacting bands with some
  quasiparticle renormalization. However, there is a clear difference
  in the way the HS gap disappears.  In Fe$_2$O$_3$
  (Fig.~\ref{fig:spec-fe2o3}) the gap can be squeezed arbitrarily
  small without qualitative changes of the spectra and the HS phase
  becomes unstable only when the gap is zero (within the numerical
  accuracy and resolution). In contrast, the transition in MnO does
  not proceed by squeezing the gap to zero, but by appearance of
  in-gap states. We use this observation below to discuss how the $d^5$ HS
  phase is destabilized.

  For the sake of clarity we simplify the local interaction to
  $\sum_{i,j,\sigma}Un_{i\sigma}n_{j-\sigma}+(U-J)n_{i\sigma}n_{j\sigma}$
  in the following discussion. We will estimate two quantities,
  the charge gap in the HS phase:
  \begin{equation}
    \label{eq:gap}
    E_\text{g}\approx E(d^6)+E(d^4)-2E(d^5)-W=U+4J-\Delta-W,
  \end{equation}
  and the difference (per atom) between the LS and HS ground states:
  \begin{equation}
    \label{eq:lstate}
    E(\text{LS})-E(\text{HS})=6J-2\Delta.
  \end{equation}
  Here $\Delta$ is the $e_g$-$t_{2g}$ crystal-field splitting, $W$ is an
  effective bandwidth, which describes reduction of the ionic gap due
  to the actual bandwidth and the charge-transfer parameter, $E(d^5)$
  is the ground-state energy of an ion in the HS state and $E(d^6)$,
  $E(d^4)$ are the energies of the corresponding electron addition and
  removal states, respectively.  The estimates are obtained by simple
  ionic considerations.  The interaction parameters $U$ and $J$ are
  considered to remain constant, while the crystal-field splitting
  $\Delta$ and the bandwidth $W$ increase with pressure.  The
  insulating HS state is destabilized when either (\ref{eq:gap}) or
  (\ref{eq:lstate}) becomes zero. Which of the two quantities reaches
  zero first depends on the parameters of the system.  We have adopted
  the terms {\it gap closing} and {\it local state transition} for the
  former and the latter instability, respectively.

  \begin{figure}
    \centering
    \includegraphics[height=0.6\columnwidth,angle=270,clip]{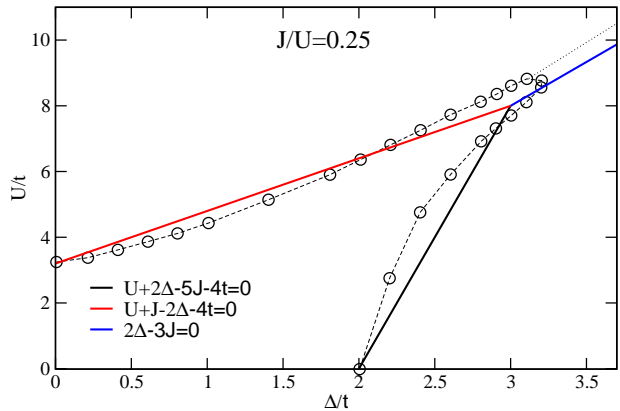}
    \caption{\label{fig:werner} Comparison of the phase boundaries of the
      two-band obtained with full DMFT (QMC) calculation (taken from Ref.~\onlinecite{werner-07})
      with the phase boundaries obtained by simple ionic estimates described in the text.
      The boundary between the ``metal'' and ``band insulator'' was obtained by setting
      the charge gap above the  band insulator ground state to zero.}
  \end{figure}

  Comparing the behavior of the charge gap in Fe$_2$O$_3$, where the HS
  phase exists as long as the gap is finite, and in MnO, where the
  transition starts while the gap is still finite, leads to the conclusion
  that the observed metal-insulator transition in Fe$_2$O$_3$ proceeds
  by the gap closing while in MnO the local state transition takes
  place.  Interestingly, a similar behavior was observed in the two-band
  model studied by Werner and Millis~\cite{werner-07}. In
  Fig.~\ref{fig:werner} we put the transition lines obtained from
  approximate expressions like (\ref{eq:gap}) and (\ref{eq:lstate})
  adapted for the two-band model on top of the numerical results
  obtained by the DMFT solution of the two-band model. Thus we can
  identify the transition between the ``Mott insulator'' and the
  ``metal'', in the language of Ref.~\onlinecite{werner-07}, with the
  gap closing and the transition between ``Mott insulator'' and the
  ``band insulator'' with the local state transition.  Note that the
  third transition (black line) exists in the two-band model with two
  electrons, but not in a five-band model with five electrons relevant
  for the studied materials.  The example of the two-band model shows
  that the gap closing and local state transition concepts are quite
  general.  Similar ionic considerations were recently
  used by Lyubutin {\it et al.}~\cite{lyu09} to other fillings of
  the $d$ shell. It was also shown that for $d^5$ Eq.~\ref{eq:gap}
  holds only if $\Delta<2J$.

  In studying the relationship between metal-insulator and spin
  transitions in Fe$_2$O$_3$ and MnO we discussed two scenarios how an
  insulating HS phase can be destabilized.  In the first one, the
  charge gap is closed due to increase of the crystal field
  as well as the bandwidths so that the system becomes metallic, at which point
  the ionic reasoning presented here and in  Ref.~\onlinecite{lyu09}
  breaks down. In the second one, the local HS becomes energetically
  unfavorable while the charge gap is still finite and the system
  transforms into a LS phase, which may be either metallic or
  insulating, e.g., LuFeO$_3$~\cite{xu01,ama09}, depending on the
  parameters of the system.

  \section{Correlation-induced atomic displacements and structural transformations}
  \label{sec:4}


  Previous applications of the LDA+DMFT approach have focused on the
  investigation of correlation effects in the electronic system with a
  given lattice structure, thereby neglecting the mutual interaction
  between electrons and ions. As a result, the influence of the
  electrons on the lattice structure was disregarded. Recent
  applications of LDA+DMFT, e.g., computations of the volume collapse
  in paramagnetic Ce~\cite{HM01,MH03,AB06} and
  Pu~\cite{SKA01,DSK03,SK04} and of the collapse of the magnetic
  moment in MnO~\cite{KL08}, did include the lattice, but only
  calculated the total energy of the correlated material as a function
  of the unit cell volume~\cite{DM09+KS09}.
  Moreover, these implementations of LDA+DMFT cannot describe the
  electronic and structural properties of electronically correlated
  materials in the case of structural transformations, e.g.,
  involving the cooperative Jahn-Teller (JT) effect~\cite{JT37,KK1,KK2,KK3,KK4}.
  This is due to the
  atomic-sphere approximation within the linearized and higher-order
  muffin-tin orbital [L(N)MTO] techniques~\cite{LNMTO1,LNMTO2}
  which are conventionally used in the
  LDA+DMFT approach. The approximation employs a spherical potential
  inside the atomic sphere, thereby neglecting multipole contributions
  to the electrostatic energy due to the distorted charge-density distribution.
  This makes the L(N)MTO technique unsuitable to determine atomic
  positions reliably.
  Instead, the recently proposed implementation of the LDA+DMFT
  approach~\cite{LB08,LK10,TL08,DK08,AL08,LG06}, which employs
  plane-wave pseudopotentials~\cite{RevPWSCF1,RevPWSCF2} and thus
  avoids the atomic sphere approximation, does not neglect such
  contributions.  Thereby it becomes possible to describe the effect
  of the distortion on the electrostatic energy~\cite{LB08,LK10}.

  Here we discuss the LDA+DMFT scheme implemented with plane-wave
  pseudopotentials~\cite{LB08,LK10,TL08} to compute structural
  transformations, e.g., structural phase stability and structure
  optimization, caused by electronic correlations.  Most importantly,
  this scheme is able to determine correlation-induced structural
  transformations in both paramagnetic and long-range ordered
  solids.  Thereby it is able to overcome the limitations of standard
  band-structure approaches and earlier implementations of the
  LDA+DMFT approach, and opens the way for fully microscopic
  investigations of the structural properties of strongly correlated
  electron materials.
  In the following we 
  present the application of this approach to the investigation of
  orbital order and the cooperative JT distortion in two prototypical
  JT materials, KCuF$_3$ and LaMnO$_3$, and compute the electronic,
  structural, and orbital properties in their room-temperature
  paramagnetic phase.
  These are the first results obtained for a structural optimization where
  the stability of the cooperative JT distortion in paramagnetic KCuF$_3$ and
  LaMnO$_3$ is investigated using total energy calculations~\cite{entropy}.

  \subsection{Application to KCuF$_3$}


  KCuF$_3$ is a prototype material regarding the cooperative
  Jahn-Teller (JT) effect, orbital ordering, and low-dimensional
  magnetism~\cite{KK1,KK2,KK3,KK4,KY67}.  It is an insulating
  pseudocubic perovskite whose structure is related to that of
  high-temperature superconductors and colossal magnetoresistance
  manganites. The copper ions have octahedral fluorine surrounding and
  are nominally in a Cu$^{2+}$ ($3d^9$) electronic configuration, with
  completely filled $t_{2g}$ orbitals and a single hole in the $e_g$
  states.
  This gives rise to a strong JT
  instability that lifts the cubic degeneracy of the Cu $e_g$ states due to a
  cooperative JT distortion~\cite{KK1,KK2,KK3,KK4}. The latter is characterized by
  CuF$_6$ octahedra elongated along the $a$ and $b$ axis and arranged
  in an antiferro-distortive pattern in the $ab$
  plane~\cite{BM90}. The strong JT distortion persists up to the melting
  temperature ($\sim$ 1000~K) and is associated with the particular
  orbital order in KCuF$_3$, in which a single hole alternatingly
  occupies $d_{x^2-z^2}$ and $d_{y^2-z^2}$ orbital states, resulting in a
  tetragonal compression ($c<a$) of the unit cell.
  Purely electronic effects as in the Kugel-Khomskii
  theory~\cite{KK1,KK2,KK3,KK4} and the electron-lattice~\cite{G63}
  interaction have been discussed as a possible mechanism behind the
  orbital ordering in KCuF$_3$.  Nevertheless, the mechanism
  responsible for the orbital order in KCuF$_3$ is still being debated
  in the
  literature~\cite{LA95,KK1,KK2,KK3,KK4,LB08,G63,MK02,PK08,BA04}.


  KCuF$_3$ has a relatively high (tetragonal, space group $I4/mcm$)
  crystal symmetry, hence it is one of the simplest systems to study.
  The JT distortion in KCuF$_3$ can be expressed using a single
  internal structure parameter, the shift of the in-plane fluorine
  atom from the Cu-Cu bond center.  Moreover, KCuF$_3$ has a single
  hole in the $3d$ shell resulting in the absence of multiplet effects.
  The electronic and structural properties of KCuF$_3$ have been
  intensively studied by density functional theory in the local
  density approximation (LDA)~\cite{LDA}, the generalized gradient
  approximation (GGA)~\cite{PB96,TB09,GGALDA}, or using the so-called
  LDA+U approach ~\cite{AZ91,LA95,AA97}.  While the LDA+U calculations
  account rather well for the value of the equilibrium JT distortion in
  KCuF$_3$~\cite{BA04}, the calculations simultaneously predict a
  long-range antiferromagnetic order which indeed occurs in KCuF$_3$
  below $T_{\mathrm{N}}$ $\sim$ 22--38~K~\cite{HS69}.
  The LDA+U calculations give the correct insulating ground state with
  the long-range $A$-type antiferromagnetic and
  $d_{x^2-z^2}/d_{y^2-z^2}$ antiferro-orbital
  order~\cite{LA95,MK02,BA04}, consistent with the
  Goodenough-Kanamori-Anderson rules for a superexchange interaction.
  Moreover, LDA+U calculations for a model structure of KCuF$_3$ in
  which cooperative JT distortions are completely neglected reproduce
  the correct orbital order, suggesting an electronic origin of the
  ordering~\cite{LA95,MK02} in agreement with the Kugel-Khomskii
  theory~\cite{KK1,KK2,KK3,KK4}.
  Nonmagnetic LDA/GGA calculations instead predict a \emph{metallic} behavior and
  cannot explain the insulating paramagnetic behavior at $T>T_{\mathrm{N}}$.
  The electronic and structural properties of KCuF$_3$ have been
  recently reexamined by means of LDA+U molecular-dynamic simulations,
  indicating a possible symmetry change and challenging the original
  assignment of tetragonal symmetry~\cite{BA04}. This symmetry change
  seems to allow for a better understanding of Raman~\cite{U91},
  electronic paramagnetic resonance~\cite{Y89,EZ08,DL08}, and x-ray
  resonant scattering~\cite{PC02,CP02} properties at $T\approx T_N$.
  However, the details of this distortion have not been fully
  understood yet.
  While the LDA+U approach is able to determine electronic properties
  and the JT distortion in KCuF$_3$ rather well~\cite{BA04}, its
  application is limited to temperatures below $T_{\mathrm{N}}$.
  Therefore LDA+U cannot explain the properties at $T>T_{\mathrm{N}}$
  and, in particular, at room temperature, where KCuF$_3$ is a
  correlated paramagnetic insulator with a robust JT distortion which
  persists up to the melting temperature. 
  To determine
  the correct orbital order and cooperative JT distortion for a
  correlated paramagnet, i.e., to perform a structural optimization,
  we here employ the novel GGA+DMFT scheme implemented with plane-wave
  pseudopotentials~\cite{LB08,LK10,TL08}.


  We start by calculating the nonmagnetic GGA band structure of
  KCuF$_3$ within the plane-wave pseudopotential
  approach~\cite{PSEUDO}.  In this calculation we use the
  Perdew-Burke-Ernzerhof exchange-correlation functional together with
  Vanderbilt ultrasoft pseudopotentials for copper and fluorine, and a
  soft Troullier-Martin pseudopotential for potassium.
  The calculation was performed for the experimental room-temperature
  crystal structure with space group $I4/mcm$ and lattice constants
  $a=5.855$ and $c=7.852$ \AA~\cite{BM90}. We have used different
  values of the JT distortion $\delta_{\mathrm{JT}}$ defined
  accordingly as $\delta_{\mathrm{JT}}=
  \frac{1}{2}(d_l-d_s)/(d_l+d_s)$;  $d_l$ and $d_s$ denote the
  long and short Cu-F bond distances in the $ab$ plane of CuF$_6$
  octahedra, respectively, and $2(d_l+d_s)=a$.  We express the
  distortion $\delta_{\mathrm{JT}}$ in percent of the lattice constant
  $a$, e.g., $\delta_{\mathrm{JT}}=0.002 \equiv 0.2$\%, and consider
  $0.2\%\leq\delta_{\mathrm{JT}} \leq7\%$. The structural
  data~\cite{BM90} at room-temperature yield $\delta_{\mathrm{JT}}=
  4.4$\%.
  In the present calculation we keep the lattice parameters $a$ and
  $c$ and the space group symmetry fixed.

  \begin{figure}
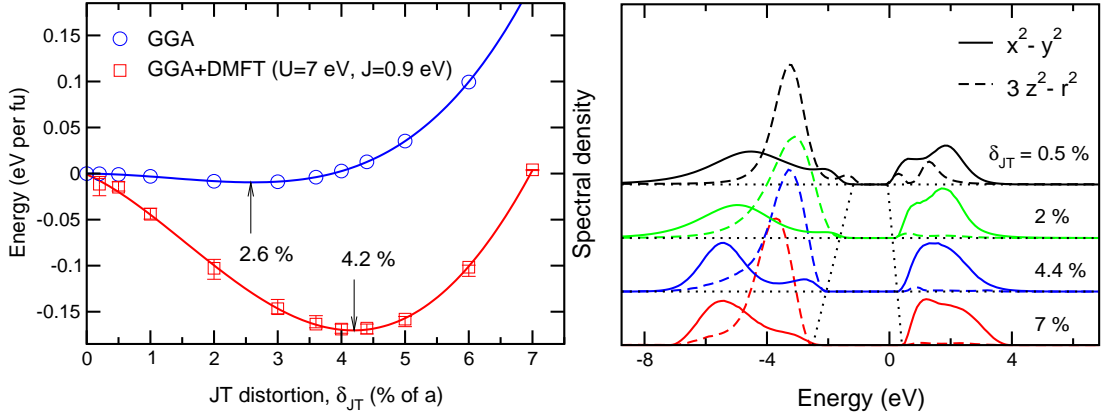

    \resizebox{\columnwidth}{!}{
      \includegraphics[clip]{KCuF3_etot.eps}
      \includegraphics[clip]{KCuF3_spectra.eps}
    }
    \caption{Left panel: Comparison of the total energies of paramagnetic
      KCuF$_3$ computed by GGA and GGA+DMFT (QMC) as a function of the
      JT distortion~\cite{LB08,LK10}.  Error bars indicate the statistical error of the
      DMFT(QMC) calculations.  Right panel: Orbitally resolved Cu $e_g$
      spectral densities of paramagnetic KCuF$_3$ as obtained by
      GGA+DMFT(QMC) for different values of the JT distortion. The
      resulting orbitally resolved spectral density which is shown
      here by solid [dashed] line is predominantly of $x^2-y^2$
      [$3z^2-r^2$] character (in the local frame
      \protect\cite{dif_orb}).}
    \label{fig:kcuf_etot}
  \end{figure}

  For all values of $\delta_{\mathrm{JT}}$ considered here, the
  nonmagnetic GGA calculations yield a \textit{metallic} rather than
  the experimentally observed insulating behavior.  We also find an
  appreciable orbital polarization due to the crystal field splitting.
  Overall, the GGA results qualitatively agree with previous
  band-structure calculations~\cite{LA95,BA04}. Thus, the
  electron-lattice interaction alone is found insufficient to
  stabilize the orbitally ordered insulating state. The GGA total
  energy profile is seen to be almost constant for
  $\delta_{\mathrm{JT}}< 4$\%, with a very shallow minimum at about
  2.6\%.  (see Fig.~\ref{fig:kcuf_etot}, left panel).  This would imply
  that KCuF$_3$ has no JT distortion for temperatures above 100~K,
  which is in clear contradiction to experiment.
  Obviously, a JT distortion by itself, without the inclusion of
  electronic correlations in the paramagnetic phase, cannot explain
  the experimentally observed orbitally ordered \emph{insulating}
  state of KCuF$_3$.


  The next step is the construction of an effective low-energy
  Hamiltonian ${\hat H_{\mathrm{GGA}}}$ for the correlated, partially
  filled Cu $e_g$ orbitals for each value of the distortion
  $\delta_{\mathrm{JT}}$ considered here. This is achieved by
  employing the pseudopotential plane-wave GGA results and making a
  projection onto atomic-centered symmetry-constrained Cu $e_g$
  Wannier orbitals~\cite{MV97,AK05,TL08}. Taking the local Coulomb
  repulsion $U=7$~eV and Hund's rule exchange $J=0.9$~eV into account,
  we obtain the following low-energy Hamiltonian for the two ($m=1,2$)
  Cu $e_g$ bands:
  \begin{eqnarray}
    {\hat H} & = & {\hat H_{\mathrm{GGA}}} + U\sum_{im} \hat n_{im\uparrow} \hat n_{im\downarrow}
    +  \sum_{i \sigma \sigma^{\prime}} (V - \delta_{\sigma
      \sigma^{\prime}} J)  \hat n_{i1\sigma}  \hat n_{i2\sigma^{\prime}} - {\hat
      H_{\mathrm{DC}}}.
    \label{eqn:hamiltn}
  \end{eqnarray}
  Here the second and third terms on the right-hand side describe the
  local Coulomb interaction between Cu $e_g$ electrons in the same and
  in different orbitals, respectively, with $V=U-2J$, and ${\hat
    H_{\mathrm{DC}}}$ is a double counting correction which accounts
  for the electronic interactions already described by the GGA (see
  below).
  We solve the many-body Hamiltonian (\ref{eqn:hamiltn}) for each
  value of $\delta_{\mathrm{JT}}$ using the single-site DMFT with
  Hirsch-Fye quantum Monte Carlo (QMC)
  calculations~\cite{HF86,full_selfconsistency}. The calculations were
  performed at $T=1160$ K ($\beta$ $=$ 10~eV$^{-1}$), using 40
  imaginary-time slices~\cite{Off-diag-elements}.
  For all values of $\delta_{\mathrm{JT}}$ considered here, we
  compute the GGA+DMFT total energy as~\cite{AB06,LB08,LK10}
  \begin{equation}
    E = E_{\mathrm{GGA}}[\rho] + \langle  \hat H_{\mathrm{GGA}} \rangle - \sum_{m,k}
    \epsilon^{\mathrm{GGA}}_{m,k}  + \langle  \hat H_{U} \rangle - E_{\mathrm{DC}},
    \label{eqn:etot}
  \end{equation}
  where $E_{\mathrm{GGA}}[\rho]$ is the total energy obtained by GGA. The third
  term on the right-hand side of Eq.~(\ref{eqn:etot}) is the sum of the
  GGA Cu $e_g$ valence-state eigenvalues and is given by the thermal average
  of the GGA Hamiltonian with the GGA Green function $G^{\mathrm{GGA}}_{\bm{k}}(i\omega_n)$:
  \begin{equation}
    \sum_{m,k} \epsilon^{\mathrm{GGA}}_{m,k} = \frac{1}{\beta}~\sum_{n,{\bm{k}}}
    \text{Tr}[H_{\mathrm{GGA}}(\bm{k}) G^{\mathrm{GGA}}_{\bm{k}}(i\omega_n)]
    e^{i\omega_n0^{+}}.
  \end{equation}
  The average $\langle \hat H_{\mathrm{GGA}} \rangle$ is evaluated
  similarly but with the full Green function including the
  self-energy. The interaction energy $\langle \hat H_{U} \rangle$ is
  computed from the double occupancy matrix. The double-counting
  correction $E_{\mathrm{DC}}= \frac{1}{2}U N_{e_g}(N_{e_g}-
  1)-\frac{1}{4}J N_{e_g}(N_{e_g}- 2)$ corresponds to the average
  Coulomb repulsion between the $N_{e_g}$ electrons in the Cu $e_g$
  Wannier orbitals. Since the Hamiltonian involves only correlated
  orbitals the number of Wannier electrons $N_{d}$ is conserved.
  Therefore, the double-counting correction amounts to an irrelevant
  shift of the total energy.


  In Fig.~\ref{fig:kcuf_etot} (left panel) we present our results for the
  GGA+DMFT total energy as a function of the JT distortion
  $\delta_{\mathrm{JT}}$. The inclusion of the electronic correlations
  among the partially filled Cu $e_g$ states in the GGA+DMFT approach
  is seen to lead to a very substantial lowering of the total energy
  by $\sim$ 175~meV per formula unit. This implies that the strong JT
  distortion persists up to the melting temperature ($>1000$ K), in
  agreement with experiment. The minimum of the GGA+DMFT total energy
  is located at the value $\delta_{\mathrm{JT}}= 4.2\%$, which is also
  in excellent agreement with the experimental value of
  4.4\%~\cite{BM90}. This clearly shows that the JT distortion in
  paramagnetic KCuF$_3$ is caused by electronic correlations.

\begin{figure}
%
%
    \centering\includegraphics[height=5cm]{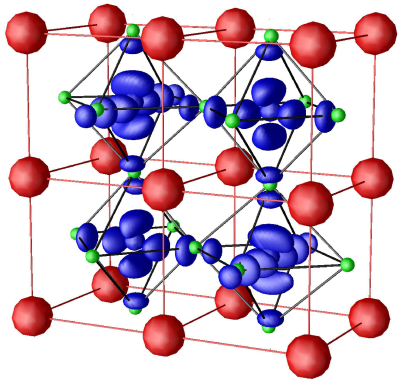}
    \caption{The $I4/mcm$ primitive cell and hole orbital order as
      obtained by the GGA+DMFT calculation for paramagnetic KCuF$_3$
      with $\protect\delta_{\mathrm{JT}}=4.4$\% is shown. The
      fluorine atoms and fluorine octahedra are shown in green, the
      potassium in red, and the Wannier Cu $e_g$ charge density in
      blue. The local coordinate system is chosen with the \emph{z}
      direction defined along the longest Cu-F bond of the CuF$_6$
      octahedron.}
    \label{fig:kcuf_wan}
  \end{figure}

  In Fig.~\ref{fig:kcuf_etot} (right panel) we show the spectral density of
  paramagnetic KCuF$_3$ calculated for several $\delta_{\mathrm{JT}}$
  values using the maximum entropy method.  The GGA+DMFT calculations
  give a paramagnetic insulating state with a substantial orbital
  polarization for all $\delta_{\mathrm{JT}}$ considered here. The
  energy gap is in the range 1.5-3.5~eV, and increases with
  increasing of $\delta_{\mathrm{JT}}$.  The sharp feature in the
  spectral density at about $-3$~eV corresponds to the fully occupied
  $3z^2-r^2$ orbital~\cite{dif_orb}, whereas the lower and upper
  Hubbard bands are predominantly of $x^2-y^2$ character and are
  located at $-5.5$~eV and 1.8~eV, respectively. The corresponding Cu
  $e_g$ Wannier charge density calculated for the experimental value
  of JT distortion of 4.4\% is presented in Fig.~\ref{fig:kcuf_wan}.
  The GGA+DMFT results clearly show an alternating occupation of the
  Cu $d_{x^2-z^2}$ and $d_{y^2-z^2}$ hole orbitals which implies
  antiferro-orbital order.
  Moreover, we have proved the stability of the paramagnetic solution at high
  temperatures (560 K) with respect to the $A$-type antiferromagnetic one.
  In this calculation we have used the experimental room-temperature crystal
  structure of KCuF$_3$ with $\delta_{\mathrm{JT}}=4.4$\% and long-range
  $A$-type antiferromagnetic order.
  However, in agreement with experiment, the calculation gives
  paramagnetic insulating solution with the orbital order as it has
  been found above.

\begin{figure}
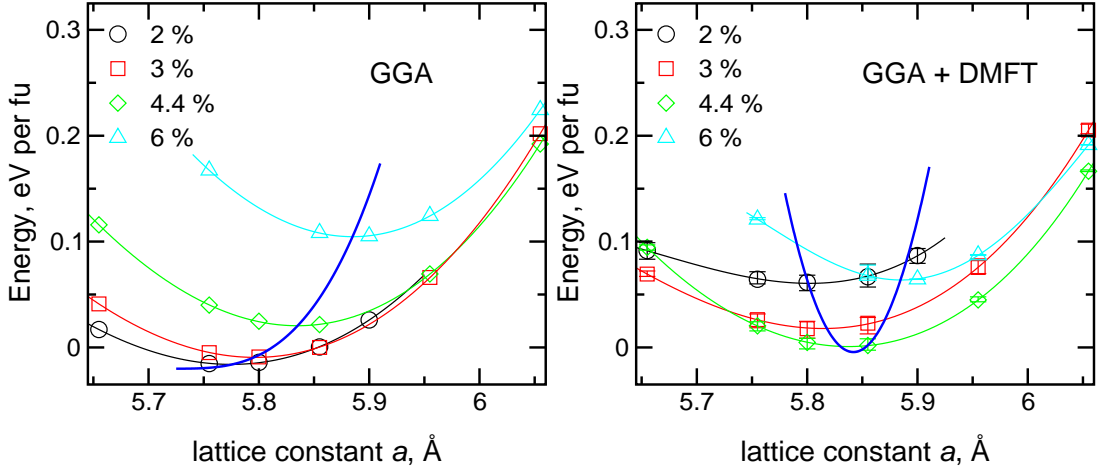

    \centering
    \resizebox{\columnwidth}{!}{
      \includegraphics[clip]{KCuF3_etot_gga.eps}
      \includegraphics[clip]{KCuF3_etot_ggadmft.eps}
    }
    \caption{Comparison of the total energies of paramagnetic
      KCuF$_3$ computed by GGA (left panel) and GGA+DMFT(QMC) (right panel) for
      different values of the JT distortion
      $\protect\delta_{\mathrm{JT}}$ as a function of the lattice
      constant~$a$~\cite{LK10}. The result of the total energy variation is 
      shown as the thick blue line.  Error bars indicate the statistical error of the
      DMFT(QMC) calculations.  }
    \label{fig:kcuf_etot_a}
  \end{figure}

  We now perform a structural optimization of paramagnetic KCuF$_3$.
  For simplicity, we consider only two independent structural
  parameters, the lattice constant $a$ and the JT distortion
  $\delta_{\mathrm{JT}}$. In this calculation we keep unchanged the
  space group symmetry (tetragonal $I4/mcm$) and the experimental
  value of the unit cell volume (taken at the ambient pressure at room
  temperature)~\cite{BM90}.
  We first calculate the non-magnetic GGA electronic structure for
  different values of $\delta_{\mathrm{JT}}$ and lattice constant $a$.
  In Fig.~\ref{fig:kcuf_etot_a} (left panel) we display the GGA total
  energies for different JT distortion $\delta_{\mathrm{JT}}$ as a
  function of the lattice constant $a$. The data points were further
  interpolated by smooth curves.
  The result of the total energy variation, namely, the line
  connecting the minima of the corresponding curves, is shown in bold
  (blue) line.  We find that the variation of the GGA total energy, is
  seen to be constant for $\delta_{\mathrm{JT}}< 2$\% with the end
  point at $a\sim 5.75$ \AA.  This implies the absence of the
  cooperative JT distortion and results in a nearly cubic ($c/a
  \approx 1.0$) unit cell, which is in clear contradiction to
  experiment~\cite{BM90}.

  To proceed further, we construct the effective low-energy
  Hamiltonian for the partially filled Cu $e_g$ orbitals for each
  value of the JT distortion $%
  \delta_{\mathrm{JT}}$ and the lattice constant $a$ considered here,
  and compute the corresponding GGA+DMFT total energy (see
  Fig.~\ref{fig:kcuf_etot_a}, right panel)~\cite{LB08,LK10}. In contrast to
  the GGA result, the inclusion of the electronic correlations among
  the partially filled Cu $e_g$ states in the GGA+DMFT method not only
  correctly describes the spectral properties, but also leads to a
  very prominent minimum in the resulting total energy variation. The
  minimum is located at the value $%
  a=5.842$ \AA\ and $\delta_{\mathrm{JT}}\approx 4.13$\%, which is in
  excellent agreement with experimental value $a=5.855$ \AA\ and
  $\delta_{\mathrm{JT}} \approx 4.4$\%. In addition, the structural
  optimization within GGA+DMFT predicts the correct tetragonal
  compression of the unit cell $c/a \approx 0.95$~\cite{BM90}.

  \subsection{Application to LaMnO$_3$}


  LaMnO$_3$ is another prototypical material with respect to the
  cooperative Jahn-Teller effect and orbital order. At
  room-temperature it is an orthorhombic perovskite with a
  GdFeO$_3$-like crystal structure and space group $Pnma$~\cite{EL71}.
  The Mn ions have octahedral oxygen surrounding and are in a
  high-spin $3d^4$ electronic configuration due to Hund's rule
  coupling, with three electrons in the $t_{2g\uparrow}$ orbitals and
  a single electron in an $e_{g\uparrow}$ orbital ($t_{2g}^3 e_g^1$
  orbital configuration). The two types of structural instabilities
  give rise to the changes relative to the cubic perovskite structure.
  The first is related to a large ion-size misfit parameter $\sqrt{2}
  (R_O + R_{Mn})/(R_O + R_{La})$ which favors rotations of the
  octahedra to accommodate a more efficient unit cell space filling;
  $R_{Mn}$, $R_{La}$, and $R_{O}$ are the ionic radii of Mn, La, and O
  ions, respectively.
  The second is a JT instability due to the $e_g$-orbital degeneracy
  which is lifted due to a cooperative JT distortion, leading to an
  alternating occupation of $d_{3x^2-r^2}$ and $d_{3y^2-r^2}$ electron
  orbitals in the $ab$ plane (antiferro-orbital ordering) and to a
  tetragonal compression of the unit cell. The rotations of the
  octahedra lower the symmetry further, finally resulting in the
  orthorhombic unit cell.
  The JT distortion experimentally persists up to $T_{\mathrm{JT}}
  \approx 750$ K which is remarkably higher than the N\'eel temperature
  of $\sim$ 140 K. Above $T_{\mathrm{JT}}$ LaMnO$_3$ undergoes a
  structural phase transition~\cite{RH98}, with volume
  collapse~\cite{CF03} to a nearly cubic structure in which orbital
  order and JT distortion vanish~\cite{RH98}.
  Below $T_N \sim$ 140 K LaMnO$_3$ shows $A$-type long-range
  antiferromagnetic order consistent with the
  Goodenough-Kanamori-Anderson rules for a superexchange interaction
  with $d_{3x^2-r^2}$/$d_{3y^2-r^2}$ antiferro-orbital
  order~\cite{EL71,RH98}.

  The LDA+U approach is found to give, at equilibrium, the correct
  insulating behavior of the low-temperature antiferromagnetic phase
  and a JT distortion in satisfactory agreement with
  experiment~\cite{TB05}. Nevertheless, it should be noted again that
  the application of this approach is limited to temperatures below
  $T_N$.  Therefore, LDA+U cannot describe the properties of LaMnO$_3$
  at $T>T_N$ and, in particular, at room temperature, where LaMnO$_3$
  is a correlated \emph{paramagnetic} insulator with a robust JT
  distortion.
  Nonmagnetic LDA/GGA calculations instead give a \emph{metal} and
  cannot explain the insulating paramagnetic behavior at $T>T_N$.
  The electronic properties of paramagnetic LaMnO$_3$ have already
  been studied using the LDA+DMFT approach~\cite{PZ00,YF06,HA08,PK09}.
  However, no attempt has been made to determine the structural
  properties and, in particular, the value of the cooperative JT
  distortion so far.  Here we investigate the stability of the
  cooperative JT distortion and perform a structural optimization of
  paramagnetic LaMnO$_3$.  To determine the electronic and structural
  properties we employ the novel GGA+DMFT scheme implemented with
  plane-wave pseudopotentials~\cite{LB08,LK10,TL08}.

  We first compute the nonmagnetic GGA band structure of paramagnetic
  LaMnO$_3$ using the plane-wave pseudopotential approach as in the
  case of KCuF$_3$.
  In this calculation, we have used the orthorhombic $Pnma$ crystal
  structure as reported by Elemans \emph{et al.}, with lattice
  constants $a=5.742$, $b=7.668$, and $c=5.532$ \AA~\cite{EL71}. The
  calculation was performed for different values of JT distortion
  $\delta_{\mathrm{JT}}$, which is now defined as the ratio between
  the difference of the long ($d_l$) and the short ($d_s$) bond
  distances and the mean Mn-O distance in the basal $ab$ plane, i.e.,
  $\delta_{\mathrm{JT}}= 2(d_l-d_s)/(d_l+d_s)$. Structural
  data~\cite{EL71} yield $\delta_{\mathrm{JT}}= 0.138$.  Here we
  change only the parameter $\delta_{\mathrm{JT}}$ $(0 \leq
  \delta_{\mathrm{JT}} \leq 0.2)$ while the value of the MnO$_6$
  octahedron tilting and rotation was fixed.

\begin{figure}
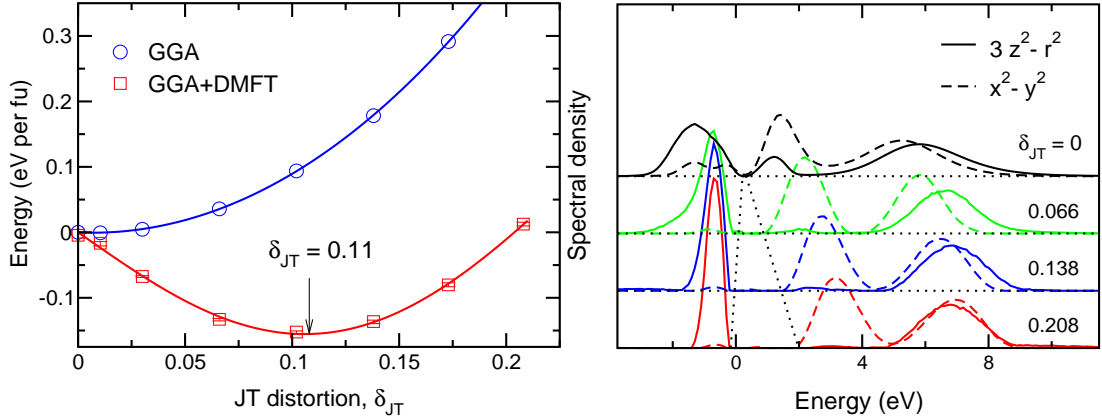

    \resizebox{\columnwidth}{!}{
      \includegraphics[clip]{LaMnO3_etot.eps}
      \includegraphics[clip]{LaMnO3_spectra.eps}
    }
    \caption{Left panel: Comparison of the total energies of paramagnetic
      LaMnO$_3$ computed by GGA and GGA+DMFT (QMC) as a function of
      the JT distortion~\cite{LK10}.  Error bars indicate the statistical error of
      the DMFT(QMC) calculations.  Right panel: Orbitally resolved Mn $e_g$
      spectral densities of paramagnetic LaMnO$_3$ as obtained by
      GGA+DMFT(QMC) for different values of the JT distortion. The
      resulting orbitally resolved spectral density shown by solid
      [dashed] line is predominantly of $3z^2-r^2$ [$x^2-y^2$]
      character (in the local frame \protect\cite{dif_orb_lamno}).  }
    \label{fig:lamno_etot}
  \end{figure}

  For all values of $\delta_{\mathrm{JT}}$ considered here we find a
  metallic solution with a considerable orbital polarization due to
  the crystal field splitting. Overall, these results qualitatively
  agree with previous band structure calculations~\cite{YV06}, namely,
  GGA cannot describe a paramagnetic insulating behavior which is
  found in experiment. We notice that even for the large
  $\delta_{\mathrm{JT}}$ value of 0.2 ($\sim 45$\% larger than found
  in experiment~\cite{EL71}) the GGA calculations predict a metal.
  In Fig.~\ref{fig:lamno_etot} (left panel) we display our results for the
  GGA total energy as a function of the JT distortion
  $\delta_{\mathrm{JT}}$.  The GGA total energy is almost parabolic
  which implies the absence of a cooperative JT distortion.
  Altogether, the non-magnetic GGA calculations give a metallic
  solution without cooperative JT distortion that contradicts
  experiment~\cite{EL71,RH98}.
  This implies the importance of electronic correlations to describe
  the experimentally observed insulating behavior and orbital order in
  paramagnetic LaMnO$_3$.

  To proceed further, we turn to the GGA+DMFT results where we treat
  the Mn $e_g$ orbitals as correlated orbitals.
  To achieve this we employ the GGA results and make a projection onto
  atomic-centered symmetry-constrained Mn $e_g$ Wannier orbitals~\cite{MV97,AK05,TL08}.
  In this calculation we assume that three (among the $3d^4$ electronic
  configuration) electrons are localized in the $t_{2g}$ orbitals. Therefore,
  they are treated as classical spins $S$, with a random orientation above
  $T_N$ (i.e., there is no correlation between different Mn sites), which
  couple to the $e_g$ electron with an energy $JS$. This coupling can be
  estimated as the energy of the splitting
  of the $e_{g \uparrow}$ and $e_{g \downarrow}$ bands in the ferromagnetic
  band-structure calculations and gives an additional term in the Hamiltonian
  (\ref{eqn:hamiltn}), namely,
  \begin{eqnarray}
    {\hat H}  =  {\hat H_{\mathrm{GGA}}} + U\sum_{im} \hat n_{im\uparrow} \hat
    n_{im\downarrow} - JS \sum_{im} (\hat n_{im\uparrow} - \hat n_{im\downarrow}
    )
    +  \sum_{i \sigma \sigma^{\prime}} (V - \delta_{\sigma
      \sigma^{\prime}} J) \hat n_{i 1\sigma} \hat n_{i 2 \sigma^{\prime}}
    - {\hat H_{\mathrm{DC}}}.
    \label{eqn:KLMhamiltn}
  \end{eqnarray}
  This corresponds to the ferromagnetic Kondo-lattice model
  Hamiltonian with an on-site Coulomb repulsion between $e_g$
  electrons, which has been intensively studied as a possible
  microscopic model to explain the colossal magnetoresistance in
  manganites~\cite{MS96,HV00}. We note that in order to calculate the
  total energy one needs to modify Eq.~\ref{eqn:etot} by adding the
  expectation value of the $JS$ term which describes the total energy
  gain due to the spin polarization of the $e_g$ orbitals at the Mn
  site.
  We take the local Coulomb repulsion $U=5$~eV, the Hund's rule
  exchange $J=0.75$~eV, and $2JS=2.7$~eV from the
  literature~\cite{YF06,HA08}, and further solve the many-body
  Hamiltonian (\ref{eqn:KLMhamiltn}) for each value of
  $\delta_{\mathrm{JT}}$ using the single-site DMFT with Hirsch-Fye
  quantum Monte Carlo (QMC)
  calculations~\cite{HF86,full_selfconsistency}. Again the calculations were
  performed at $T=1160$ K ($\beta$ $=$ 10~eV$^{-1}$), using 40
  imaginary-time slices~\cite{Off-diag-elements}.


  In Fig.~\ref{fig:lamno_etot} (left panel) we display the GGA+DMFT total
  energy calculated as a function of the JT distortion
  $\delta_{\mathrm{JT}}$. In contrast to the GGA result, GGA+DMFT
  yields a substantial total energy gain of $\sim$ 150~meV per formula
  unit. This implies that the cooperative JT distortion indeed
  persists up to high temperatures ($T>1000$ K), while in GGA a JT
  distortion does not occur at all. Taking into account that the
  calculations have been performed for the low-temperature crystal
  structure of LaMnO$_3$~\cite{EL71} this estimate (150~meV) is in
  good agreement with $T_{\mathrm{JT}} \sim 750$~K at which the JT
  distortion vanishes~\cite{RH98,CF03}.
  The minimum of the GGA+DMFT total energy is located at the value
  $\delta_{\mathrm{JT}} \sim 0.11$, which is also in good agreement
  with the experimental value of 0.138~\cite{EL71,RH98}. We note that
  GGA+DMFT calculations correctly describe both electronic and
  structural properties of paramagnetic LaMnO$_3$. This shows that the
  JT distortion in paramagnetic LaMnO$_3$ is caused by electronic
  correlations.

\begin{figure}
%
%
    \centering\includegraphics[height=5cm]{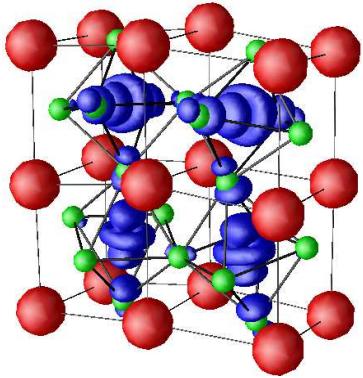}    
    \caption{The $Pnma$ primitive cell and orbital order as obtained
      by the GGA+DMFT calculation for paramagnetic LaMnO$_3$ with
      $\protect\delta_{\mathrm{JT}}=0.138$ is shown. The oxygen atoms
      and oxygen octahedra are shown in green, the lanthanum in red,
      and the Wannier Mn $e_g$ charge density in blue.  }
    \label{fig:lamno_wan}
  \end{figure}

  In Fig.~\ref{fig:lamno_etot} (right panel) we display the spectral density
  of paramagnetic LaMnO$_3$ calculated for several
  $\delta_{\mathrm{JT}}$ values using the maximum entropy method. We
  find a strong orbital polarization for large $\delta_{\mathrm{JT}}$,
  which gradually decreases for decreasing JT distortion. The occupied
  part of the $e_g$ density is located at about $-2$ to $-1$~eV and
  corresponds to the $e_g$ states with spin parallel to the $t_{2g}$
  spin at that site. It has predominantly Mn $d_{3x^2-r^2}$ and
  $d_{3y^2-r^2}$ orbital character with a considerable admixture of
  $d_{z^2-r^2}$ for small JT distortions. The energy gap is about 2~eV
  for large $\delta_{\mathrm{JT}}$ and considerably decreases with
  decreasing $\delta_{\mathrm{JT}}$, resulting in a pseudogap behavior
  at the Fermi level for $\delta_{\mathrm{JT}}=0$.
  In Fig.~\ref{fig:lamno_wan} we present the corresponding Mn $e_g$
  Wannier charge density computed for the experimental JT distortion
  value of $\delta_{\mathrm{JT}}=0.138$. The result clearly indicates
  an alternating occupation of the Mn $d_{3x^2-r^2}$ and
  $d_{3y^2-r^2}$ orbitals, corresponding to the occupation of a
  $3z^2-r^2$ orbital in the local frame~\cite{dif_orb_lamno}, which
  implies antiferro-orbital order. Thus, in agreement with experiment,
  the calculations give a paramagnetic insulating solution with
  antiferro-orbital order and stable JT distortion.

  To conclude, we presented applications of the novel GGA+DMFT
  approach to two prototypical Jahn-Teller materials, KCuF$_3$ and
  LaMnO$_3$, and computed the orbital order and cooperative JT
  distortion in these compounds.  In particular, our results obtained
  for the paramagnetic phase of KCuF$_3$ and LaMnO$_3$, namely an
  equilibrium Jahn-Teller distortion $\delta_{\mathrm{JT}}$ of 4.2\%
  and 0.11, respectively, and antiferro-orbital order, agree well with
  experiment.  The present approach overcomes the limitations of the
  LDA+U method and previous implementations of LDA+DMFT and is able to
  determine correlation-induced structural transformations in both
  paramagnetic and long-range magnetically ordered solids.  Thereby it
  can be employed for the lattice optimization and molecular-dynamic
  simulations of these systems.
  The computational scheme presented here allows us to investigate
  complex materials with strong electronic interactions and
  correlation-induced structural transformations, shifts of
  equilibrium atomic positions and changes in the lattice structure,
  and to perform a structural optimization of \emph{paramagnetic}
  solids.
  It opens the way for fully microscopic investigations of the
  structural properties of strongly correlated electron materials such
  as lattice instabilities observed at correlation-induced
  metal-insulator transitions.

  \section{Kinks in the dispersion of correlated materials}
  \label{sec:5}

  The electronic dispersion relation indicates at which energy and crystal
  momentum one-particle excitations can occur in a solid. Interactions between
  electrons influence not only the shape and position of the
  dispersion, but also make the lifetime of such excitations finite.
  A robust general definition for the branches of the dispersion
  relation $E_{\bm{k}}$ is provided by the local maxima of the spectral
  function $A({\bm{k}},\omega)$, i.e., its peak locations as a
  function of $\omega$ for a given momentum $\bm{k}$. For energies
  $\omega$ far from the Fermi surface, the peaks in the spectral
  function of a strongly interacting material can be quite broad,
  corresponding to incoherent excitations with a short lifetime.
  Nevertheless the maxima in $A({\bm{k}},\omega)$ are always well
  defined, even if their location might be difficult to determine
  from angle-resolved photoemission data at high binding energies.
  The excitations for energies close to the Fermi surface, on the
  other hand, are coherent in a Landau-Fermi liquid, i.e., they are only
  weakly damped and correspond to sharp peaks in the spectral
  function.  In this case the correlated dispersion is given by the
  uncorrelated dispersion multiplied by a renormalization factor $Z$
  related to Fermi-liquid parameters.

\begin{figure}
    \centering\resizebox{0.65\columnwidth}{!}{
      \includegraphics[clip]{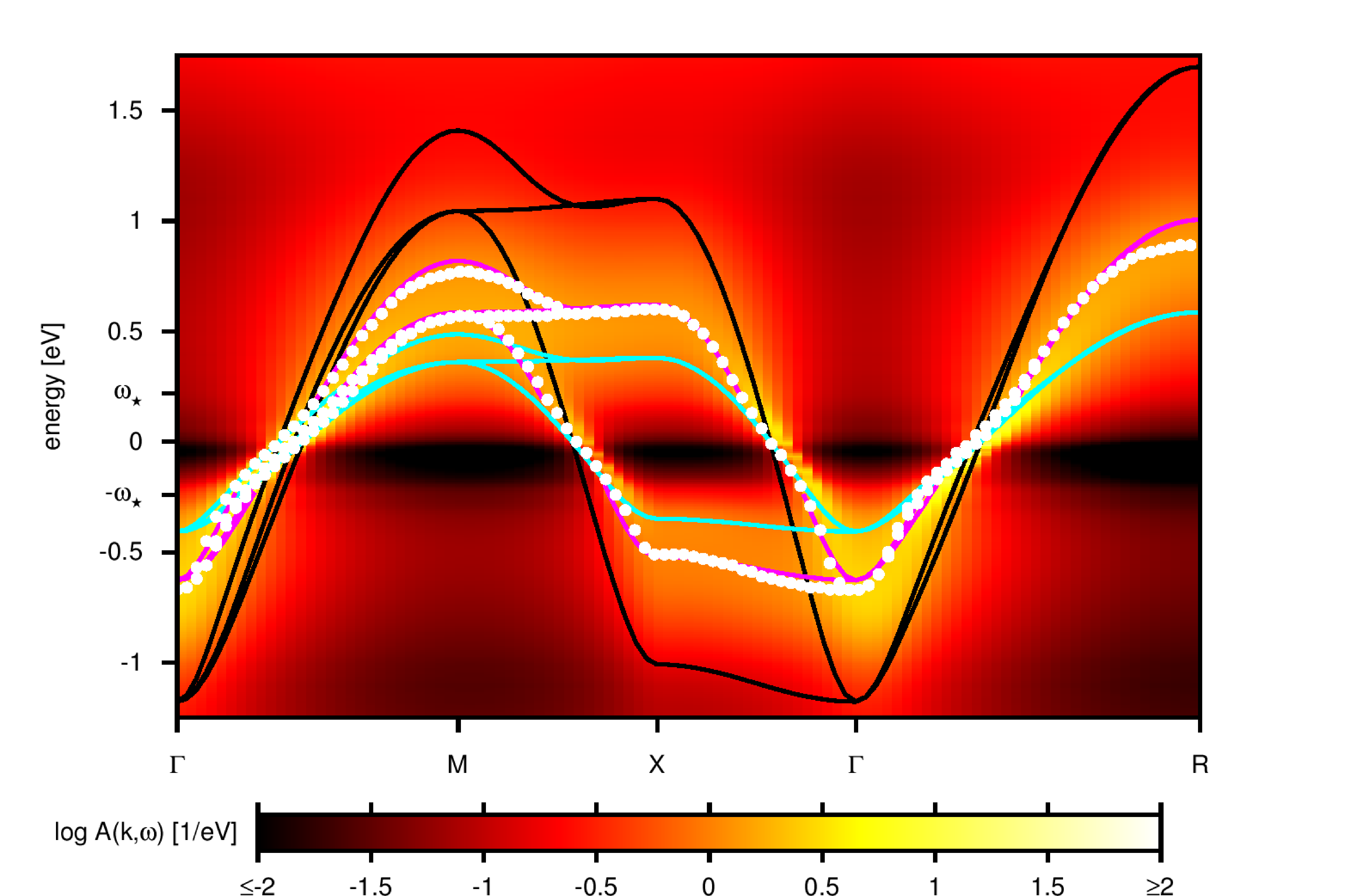}
    }
    \caption{Intensity plot of the total spectral function for
      SrVO$_3$ obtained with LDA+DMFT; after~\cite{E6:Nekrasov2006a}. Near
      the Fermi energy the dispersion $E_{\bm{k}}$ (white dots) is
      given by the renormalized LDA band structure
      $\epsilon_{\bm{k}}$ (black lines) with a Fermi-liquid factor
      $Z$ $=$ $0.35$, i.e.  $E_{\bm{k}}$ $=$ $Z\epsilon_{\bm{k}}$ (blue
      line).  For higher energies the dispersion relation follows the
      LDA band structure with a different renormalization,
      $E_{\bm{k}}$ $=$ $Z'\epsilon_{\bm{k}}+ c_\pm$ (pink line), with
      $Z'$ $=$ $0.64$, $c_+$ $=$ 0.086~eV, $c_-$ $=$ 0.13~eV \cite{E6:Byczuk07a}.
      The crossover between the two slopes leads to kinks at energies
      $\omega_{\star,+}$ $=$ 0.22~eV and $\omega_{\star,-}$ $=$ -0.24~eV in
      the effective dispersion. The parameter values given here are
      calculated from the theory of Ref.~\cite{E6:Byczuk07a}; due to
      the shift $c_\pm$ the parameter $Z'$ differs from the fitted
      parameter in Ref.~\cite{E6:Nekrasov2006a} without shift.
      \label{fig:srvo3}}
  \end{figure}

  A characteristic feature of strongly correlated metals is a transfer
  of spectral weight into the lower and upper Hubbard bands due to the
  Coulomb interaction.  Together with the central peak near the Fermi
  energy the momentum-integrated spectral function therefore typically
  has three peaks. One exemplary material with this type of spectrum
  is the cubic perovskite SrVO$_3$ with $3d^1$ electronic
  configuration, leading to band-filling of one electron per site in
  the threefold-degenerate $t_{2g}$ bands. To obtain its
  momentum-resolved spectral function with LDA+DMFT, we used $N$th
  order muffin orbitals (NMTO) to obtain the downfolded $3\times3$
  Hamiltonian for the $t_{2g}$ bands~\cite{E6:Nekrasov2006a}.  For the
  DMFT calculation the Coulomb interaction parameters for these bands
  are $U$ $=$ 5.55~eV for the on-site repulsion, $J$ $=$ $1.0$~eV for
  the Hund's exchange coupling, and $U'$ $=$ 3.55~eV for the
  interorbital Coulomb repulsion~\cite{Gunnarsson89}.  The local
  (momentum-integrated) spectral function has an asymmetric three-peak
  structure with a high central peak and two smaller peaks due to the
  Hubbard bands~\cite{E6:Nekrasov2006a}.  The total spectral
  function $A({\bm{k}},\omega)$ is shown in the intensity plot in
  Fig.~\ref{fig:srvo3},
  with its local maxima marked by white dots.  Remarkably, the
  dispersion $E_{\bm{k}}$ shows a small sudden change in slope near
  energies $\omega_{\star}\approx0.25$~eV. There a crossover between
  two regimes occurs: In the low-energy Fermi-liquid regime,
  $|\omega|$ $<$ $\omega_{\star}$, the LDA dispersion
  $\epsilon_{\bm{k}}$ is renormalized by the Landau parameter $Z$
  so that $E_{\bm{k}}=Z\epsilon_{\bm{k}}$. On the other hand, at
  higher energies, outside the Fermi-liquid regime but still inside
  the central peak, the dispersion is given by
  $E_{\bm{k}}=Z'\epsilon_{\bm{k}}$ with a different
  renormalization factor. In Ref.~\cite{E6:Nekrasov2006a} these
  parameters were determined as $Z=0.33$ and $Z'=0.53$, i.e., the
  renormalization is much stronger near the Fermi surface. This value
  of $Z'\approx0.5$ was also found in ARPES experiments
  \cite{Yoshida05,Takizawa09}. Although no sharp kink was found, a
  weak change in slope was detected at
  $\omega\approx-0.2$~eV~\cite{Takizawa09}, in agreement with the
  LDA+DMFT calculation.

\begin{figure}
    \centering\resizebox{0.55\columnwidth}{!}{
      \includegraphics[clip]{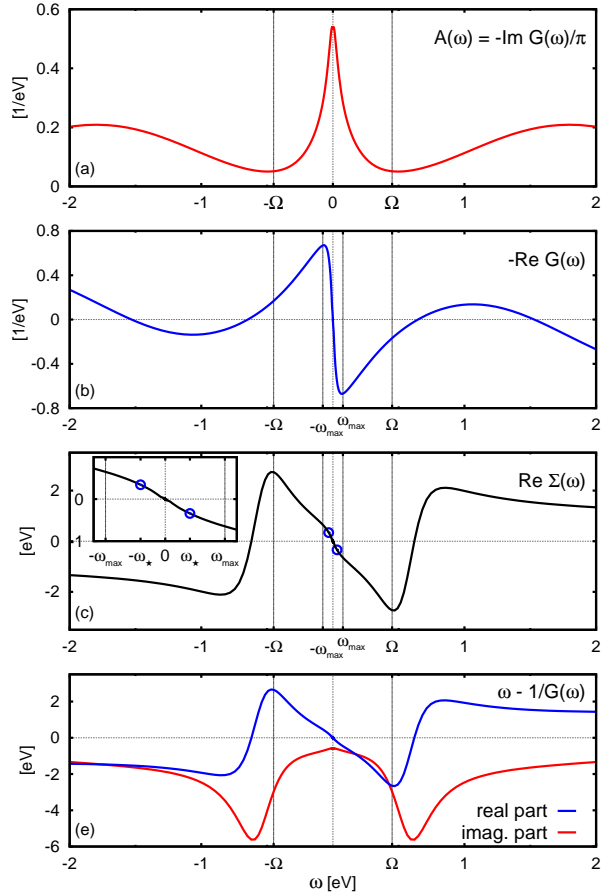}
    }
    \caption{Origin of kinks in a strongly correlated system (one-band
      Hubbard model in DMFT with $Z=0.086$)~\cite{E6:Byczuk07a}.
      From top to bottom:
      Local spectral function with three well-separated peaks;
      corresponding real part of the local Green function with peaks
      inside central spectral peak; real part of the self-energy, in
      which the maxima of $\text{Re}[G(\omega)]$ translate into kinks;
      $\omega-1/G(\omega)$, which contributes only a smooth linear
      background to $\text{Re}[\Sigma(\omega)]$.}
    \label{fig:kinks}
  \end{figure}

  These kinks in the dispersion can be understood from a microscopic
  theory \cite{E6:Byczuk07a}, which correctly predicts the
  renormalized bandstructure outside the Fermi-liquid regime only from
  Fermi-liquid parameters and the uncorrelated spectrum.  A
  sufficiently strong interaction is needed, so that the local
  spectral function has a three-peak structure with a central peak and
  two Hubbard peaks. In the absence of particle-hole symmetry this
  condition can be relaxed; e.g., it suffices that the central peak is
  well separated from the lower Hubbard peak for a kink to develop
  below the Fermi surface. As we will now discuss, the dips on the
  sides of the central peak lead to kinks in the dispersion inside the
  central peak; for simplicity, we consider only a one-band system
  with symmetric spectrum (Fig.~\ref{fig:kinks}).
  Namely, well-separated peaks in the local spectral function
  $A(\omega)=-\text{Im}[G(\omega)/\pi]$ (top panel in
  Fig.~\ref{fig:kinks}) have the consequence that due to
  Kramers-Kronig relations, $\text{Re}[G(\omega)]$ must have sharp
  maxima at energies $\omega_{\text{max}}$ inside the central spectral
  peak (second panel from top in Fig.~\ref{fig:kinks}). These maxima
  induce kinks in the self-energy $\Sigma(\bm{k},\omega)$, which lead
  to kinks in the $\bm{k}$-resolved spectral function,
  $A(\bm{k},\omega)$ $=$
  $-\text{Im}[\omega+\mu-\epsilon_{\bm{k}}-\Sigma(\bm{k},\omega)]^{-1}/\pi$,
  and thus in the dispersion.  In DMFT the self-energy is independent
  of momentum, $\Sigma(\bm{k},\omega)=\Sigma(\omega)$, and related to
  the local Green function by the self-consistency condition,
  $\Sigma(\omega)=\omega+\mu-1/G(\omega)-\Delta(G(\omega))$, where
  $\Delta(G(\omega))$ is the hybridization function. For a three-peak
  spectral function $A(\omega)$, the term
  $\text{Re}[\omega+\mu-1/G(\omega)]$ is essentially linear inside the
  central peak (bottom panel in Fig.~\ref{fig:kinks}). By contrast,
  $\Delta(G(\omega))$, which can be expanded in powers of $G(\omega)$,
  always contains a term linear in $G(\omega)$. As a consequence
  $\Sigma(\omega)$ always has a contribution from the peaks in
  $\text{Re}[G(\omega)]$ that is nonlinear in $\omega$ inside the
  central peak. Thus $\text{Re}[\Sigma(\omega)]$ has two different
  slopes for $|\omega|$ smaller or larger than $\omega_{\star}$.  This
  translates into the usual Fermi-liquid renormalization
  $E_{\bm{k}}=Z\epsilon_{\bm{k}}$ near the Fermi energy, but a shifted
  and renormalized dispersion $E_{\bm{k}}=Z'\epsilon_{\bm{k}}$ $+$
  const for higher binding energies~\cite{E6:Byczuk07a}. From the
  Fermi-liquid renormalization $Z$ and the non-interacting
  bandstructure the constant shift, the parameter $Z'$ and the
  crossover energy $\omega_{\star}$ can be obtained
  explicitly~\cite{E6:Byczuk07a}; in particular $Z'$ and
  $\omega_{\star}$ are proportional to the Fermi-liquid parameter $Z$
  (multiplied by an uncorrelated scale).  The values for SrVO$_3$ are
  listed in Fig.~\ref{fig:srvo3} and fit the correlated dispersion very
  well.

  According to the theory described above the kinks in the slope of
  the dispersion are a direct consequence of the electronic
  interaction~\cite{E6:Byczuk07a}. They can lead to corresponding
  kinks in the low-temperature electronic specific
  heat~\cite{Toschi2009}. The presence of kinks has also been linked
  to maxima in the spin susceptibility~\cite{E6:Uhrig2009}. Of course,
  additional kinks in the electronic dispersion may arise from the
  coupling of electrons to bosonic degrees of freedom, such as
  phonons.  Recent experiments have found evidence for kinks in
  Ni(110)~\cite{E6:Nickel2009}, which may be due to the electronic
  mechanism presented here.

  \section{Summary}

  The combination of dynamical mean-field theory (DMFT) with LDA/GGA
  electronic structure calculations has led to important advances in
  our understanding and in the quantitative description of correlated
  electron materials. For example, by including the $p$-$d$
  hybridization into the LDA+DMFT scheme it is now possible to clearly
  distinguish between Mott-Hubbard and charge-transfer insulators.
  This allowed for the computation of the local correlations and
  hole doping in NiO. By treating the local correlations and the
  Ni 3$d$ - O2 $p$ hybridization on the same footing we were able to
  provide a description of the full valence-band and conduction-band
  spectra of charge-transfer systems with strong hybridization.
  These results provide a quantitative explanation of the experimental
  photoemission, inverse photoemission, and ARPES data. 
  Our results show that, provided the $p$-$d$
  hybridization is explicitly included, the LDA+DMFT is able to treat
  the late transition-metal oxides and charge-transfer systems in
  general.

  In a second application we investigated the metallization and
  magnetic-moment collapse in MnO and Fe$_2$O$_3$. Our results not
  only describe the experimentally observed simultaneous collapse of
  the magnetic moment and the volume in MnO but also explain the
  origin of the collapse of the magnetic moment which originates from
  the increase of the crystal field splitting and not from the change
  of the band width. Since similar results were obtained for hematite
  (Fe$_2$O$_3$) we may conclude that the transition in this compound
  at about 50 GPa may be described by an electronically driven volume
  collapse.

  Our formulation of GGA+DMFT in terms of plane-wave pseudopotentials
  provides a robust computational scheme for the investigation of
  complex correlated materials. 
We applied this novel approach to explain various
electronic and structural properties of the two prototypical Jahn-Teller
materials KCuF$_3$ and LaMnO$_3$ in their correlated paramagnetic phase. Our results for the equilibrium cooperative JT distortion, the
lattice constants, and the antiferro-orbital order agree well with experiment.
This computational scheme will make it possible to investigate 
correlation induced structural transformations,
shifts of equilibrium atomic positions, and changes in the lattice structure,
and to perform the structural optimization of correlated \emph{paramagnetic} solids.
This opens the way for fully
  microscopic investigations of lattice instabilities which are known
  to occur in the vicinity of correlation-induced metal-insulator
  transitions.

  Finally, we offered a comprehensive explanation of kinks in the
  effective dispersion of correlated electrons in correlated materials
  with a pronounced three-peak structure. In particular we found that
  Landau quasiparticles exist only in a remarkably narrow region
  around the Fermi energy. For higher energies, but still in the
  central maximum of the spectral function, the excitations are
  described by a different dispersion relation which may be calculated
  analytically. The transition from one to the other regime is
  relatively sharp and leads to the kinks in the dispersion relation.
  One of the most astonishing results of this investigation is the
  fact that it is possible to describe properties of the system
  outside the Landau-Fermi-liquid regime fully analytically.

  In view of the rapid developments of the LDA+DMFT scheme during the
  last couple of years it is clear that this approach has a great
  potential for further developments. It is already foreseeable that
  the LDA+DMFT approach will even be able to explain and predict the
  properties of complex correlated electron materials.

  \medskip
  {\small
  We thank M. Altarelli, N. Binggeli, R. Bulla, J. Deisenhofer, J.
  Fink, A. Fujimori, D. Khomskii, Dm. Korotin, S. Streltsov, and G. Trimarchi for
  useful discussions. Financial support by the SFB 484 and TTR 80 of
  the Deutsche Forschungsgemeinschaft is gratefully acknowledged.
  }


\begin{thebibliography}{999}


  \bibitem{Rev1} M. Imada, A. Fujimori, Y. Tokura, Rev. Mod. Phys.
    \textbf{70}, 1039 (1998)

  \bibitem{Rev2} Y. Tokura, N. Nagaosa, Science \textbf{288}, 462
    (2000)

  \bibitem{Rev3} E. Dagotto, Science \textbf{309}, 257 (2005)

  \bibitem{LDA} R. O. Jones, O. Gunnarsson, Rev. Mod. Phys.
    \textbf{61}, 689 (1989)

  \bibitem{PB96} J. P. Perdew, K. Burke, M. Ernzerhof, Phys. Rev.
    Lett.  \textbf{77}, 3865 (1996)

  \bibitem{TB09} F. Tran, P. Blaha, Phys. Rev. Lett. \textbf{102},
    226401 (2009)

  \bibitem{AZ91} V. I. Anisimov, J. Zaanen, O. K. Andersen, Phys.
    Rev. B \textbf{44}, 943 (1991)

  \bibitem{LA95} A. I. Liechtenstein, V. I. Anisimov, J. Zaanen,
    Phys. Rev. B \textbf{52}, R5467 (1995)

  \bibitem{AA97} V. I. Anisimov, F. Aryasetiawan, A. I.
    Lichtenstein, J. Phys.: Condens. Matter \textbf{9}, 767 (1997)


  \bibitem{RevPWSCF1} S. Baroni, S. de Gironcoli, A. Dal Corso, P.
    Giannozzi, Rev. Mod. Phys. \textbf{73}, 515 (2001)

  \bibitem{RevPWSCF2} P. Giannozzi, S. Baroni, N. Bonini, M. Calandra,
    R. Car, C. Cavazzoni, D. Ceresoli, G. L. Chiarotti, M. Cococcioni,
    I. Dabo, A. Dal Corso, S. Fabris, G. Fratesi, S. de Gironcoli, R.
    Gebauer, U. Gerstmann, C. Gougoussis, A. Kokalj, M. Lazzeri, L.
    Martin-Samos, N. Marzari, F. Mauri, R. Mazzarello, S. Paolini, A.
    Pasquarello, L. Paulatto, C. Sbraccia, S. Scandolo, G. Sclauzero,
    A. P. Seitsonen, A. Smogunov, P. Umari, R. M. Wentzcovitch, J.
    Phys. Condens. Matter \textbf{21}, 395502 (2009)

  \bibitem{RevWIEN2k} P. Blaha, K. Schwarz, G. Madsen, D. Kvasnicka,
   J.  Luitz, WIEN2k, \emph{An Augmented Plane Wave + Local
      Orbitals Program for Calculating Crystal Properties} (Karlheinz
    Schwarz, Techn. Universit{\"a}t Wien, Austria), 2001.

  \bibitem{RevVASP} G. Kresse, J. Hafner, Phys. Rev. B \textbf{47},
    558 (1993)

  \bibitem{DMFT1} W. Metzner, D. Vollhardt, Phys. Rev. Lett.
    \textbf{62}, 324 (1989)

  \bibitem{DMFT2} A. Georges, G. Kotliar, W. Krauth, M. J.
    Rozenberg, Rev. Mod. Phys. \textbf{68}, 13 (1996)

  \bibitem{DMFT3} G. Kotliar, D. Vollhardt, Phys. Today
    \textbf{57}, 53 (2004)

  \bibitem{DMFT4} G. Kotliar, S. Y. Savrasov, K. Haule, V. S.
    Oudovenko, O. Parcollet, C. A. Marianetti, Rev. Mod. Phys.
    \textbf{78}, 865 (2006)

  \bibitem{DMFT5} M. I. Katsnelson, V. Yu. Irkhin, L. Chioncel, A. I.
    Lichtenstein, R. A. de Groot, Rev. Mod. Phys. 80, 315 (2008)

  \bibitem{DMFTmeth1} V. I. Anisimov, A. I. Poteryaev, M. A. Korotin,
    A. O. Anokhin, G. Kotliar, J. Phys. Condens. Matt. \textbf{9},
    7359 (1997)

  \bibitem{DMFTmeth2} A. I. Lichtenstein, M. I. Katsnelson, Phys.
    Rev. B \textbf{57}, 6884 (1998)

  \bibitem{DMFTmeth3} A. I. Lichtenstein, M. I. Katsnelson, G.
    Kotliar, in \emph{Electron Correlations and Materials Properties
      2nd ed.}, edited by A. Gonis, N. Kioussis, M. Ciftan (Kluwer
    Academic/Plenum, New York, 2002), p. 428

  \bibitem{DMFTmeth4} K. Held, I. A. Nekrasov, G. Keller, V. Eyert, N.
    Bl{\"u}mer, A. K. McMahan, R. T. Scalettar, Th. Pruschke, V. I.
    Anisimov, D. Vollhardt, Psi-k Newsletter \textbf{56}, 65
    (2003);

  \bibitem{DMFTmeth5} K. Held, I. A. Nekrasov, G. Keller, V. Eyert, N.
    Bl{\"u}mer, A. K. McMahan, R. T. Scalettar, Th. Pruschke, V. I.
    Anisimov, D. Vollhardt, Phys. Status Solidi B \textbf{243},
    2599 (2006)

  \bibitem{DMFTmeth6} K. Held, Adv. Phys. 56, 829 (2007)

  \bibitem{DMFTcalc1} K. Held, G. Keller, V. Eyert, D. Vollhardt,
    V. I. Anisimov, Phys. Rev. Lett. \textbf{86}, 5345 (2001)

  \bibitem{DMFTcalc2} E. Pavarini, S. Biermann, A. Poteryaev, A. I.
    Lichtenstein, A. Georges, O. K. Andersen, Phys. Rev. Lett.
    \textbf{92}, 176403 (2004)

  \bibitem{DMFTcalc3} A. I. Poteryaev, A. I. Lichtenstein, G.
    Kotliar, Phys. Rev. Lett. \textbf{93}, 086401 (2004)

  \bibitem{DMFTcalc4} S. Biermann, A. Poteryaev, A. I. Lichtenstein,
    A. Georges, Phys. Rev. Lett.  \textbf{94}, 026404 (2005)

  \bibitem{DMFTcalc5} L. Chioncel, Ph. Mavropoulos, M. Lezai\' c, S.
    Bl{\"u}gel, E. Arrigoni, M. I. Katsnelson, A. I. Lichtenstein,
    Phys. Rev. Lett. \textbf{96}, 197203 (2006)


  \bibitem{HM01} K. Held, A. K. McMahan, R. T. Scalettar, Phys.
    Rev. Lett.  \textbf{87}, 276404 (2001)

  \bibitem{MH03} A. K. McMahan, K. Held, R. T. Scalettar, Phys.
    Rev. B \textbf{67}, 075108 (2003)

  \bibitem{AB06} B. Amadon, S. Biermann, A. Georges, F.
    Aryasetiawan, Phys. Rev. Lett. \textbf{96}, 066402 (2006); see
    also L. V. Pourovskii, B. Amadon, S. Biermann, A. Georges,
    Phys. Rev. B \textbf{76}, 235101 (2007) where the authors discuss
    the problem of full self-consistency over the charge density in
    $\gamma$-Ce and Ce$_2$O$_3$.

  \bibitem{SKA01} S. Y. Savrasov, G. Kotliar, E. Abrahams, Nature
    (London) \textbf{410}, 793 (2001)

  \bibitem{DSK03} X. Dai, S. Y. Savrasov, G. Kotliar, A. Migliori, H.
    Ledbetter, E. Abrahams, Science \textbf{300}, 953 (2003)

  \bibitem{SK04} S. Y. Savrasov, G. Kotliar, Phys. Rev. B
    \textbf{69}, 245101 (2004)

  \bibitem{DMFTcalc+1} M. I. Katsnelson, A. I. Lichtenstein Phys.
    Rev. B \textbf{61}, 8906 (2000)

  \bibitem{DMFTcalc+2} A. I. Lichtenstein, M. I. Katsnelson, G.
    Kotliar, Phys. Rev. Lett. \textbf{87}, 067205 (2001)

  \bibitem{DMFTcalc+3} J. Braun, J. Min\' ar, H. Ebert, M. I.
    Katsnelson, A. I. Lichtenstein, Phys. Rev. Lett. \textbf{97},
    227601 (2006)

  \bibitem{DMFTcalc+4} A. Grechnev, I. Di Marco, M. I. Katsnelson, A.
    I. Lichtenstein, J. Wills, O. Eriksson, Phys. Rev. B
    \textbf{76}, 035107 (2007)

  \bibitem{DMFTcalc+5} S. Chadov, J. Min\'{a}r, M. I. Katsnelson, H.
    Ebert, D. K\"{o}dderitzsch, A. I. Lichtenstein, Europhys.
    Lett. \textbf{82}, 37001 (2008)

  \bibitem{kun07a} J. Kune\v{s}, V. I. Anisimov, A. V. Lukoyanov,
    D. Vollhardt, Phys. Rev. B \textbf{75}, 165115 (2007)

  \bibitem{kun07b} J. Kune\v{s}, V. I. Anisimov, S. L. Skornyakov, A.
    V. Lukoyanov, D. Vollhardt, Phys. Rev. Lett. \textbf{99},
    156404 (2007)

  \bibitem{KL08} J. Kune\v{s}, A. V. Lukoyanov, V. I. Anisimov, R. T.
    Scalettar, W. E. Pickett, Nature Materials {\bf 7}, 198 (2008)

  \bibitem{fe2o3} J. Kune\v{s}, Dm.~M. Korotin, M.~A. Korotin, V.~I.
    Anisimov,  P. Werner, Phys. Rev. Lett.  {\bf 102}, 146402 (2009)

  \bibitem{ZSA85} J. Zaanen, G. A. Sawatzky, J. W. Allen, Phys.
    Rev.  Lett. \textbf{55}, 418 (1985)

  \bibitem{saw84} {{G.~A.} {Sawatzky}}, {{J.~W.} {Allen}}, {Phys.
      Rev. Lett.} \textbf{{53}}, {2339} ({1984})


  \bibitem{eastman} {{D.~E.} {Eastman}},{{J.~L.} {Freeouf}},
    {Phys. Rev. Lett.} \textbf{{34}}, {395} ({1975})

  \bibitem{she90} {{Z.-X.} {Shen}}, {{C.~K.} {Shih}}, {{O.}~{Jepsen}},
    {{W.~E.} {Spicer}}, {{I.}~{Lindau}}, {{J.~W.} {Allen}},
    {Phys. Rev. Lett.} \textbf{{64}}, {2442} ({1990})

  \bibitem{she91} {{Z.-X.} {Shen}}, {{R.~S.} {List}}, {{D.~S.}
      {Dessau}}, {{B.~O.} {Wells}}, {{O.}~{Jepsen}}, {{A.~J.} {Arko}},
    {{R.}~{Barttlet}}, {{C.~K.} {Shih}}, {{F.}~{Parmigiani}}, {{J.~C.}
      {Huang}}, {{P.~A.~P.} {Lindberg}}, {Phys. Rev. B}
    \textbf{{44}}, {3604} ({1991})

  \bibitem{tje96} {{O.}~{Tjernberg}}, {{S.}~{S\"oderholm}},
    {{G.}~{Chiaia}}, {{R.}~{Girard}}, {{U.~O.} {Karlsson}},
    {{H.}~{Nyl\'en}}, {{I.}~{Lindau}}, {Phys. Rev. B}
    \textbf{{54}}, {10245} ({1996})

  \bibitem{MV97} N. Marzari, D. Vanderbilt, Phys. Rev. B
    \textbf{56}, 12847 (1997)

  \bibitem{AK05} V. I. Anisimov, D. E. Kondakov, A. V. Kozhevnikov, I.
    A. Nekrasov, Z. V. Pchelkina, J. W. Allen, S.-K. Mo, H.-D. Kim, P.
    Metcalf, S. Suga, A. Sekiyama, G. Keller, I. Leonov, X. Ren,
    D. Vollhardt, Phys. Rev. B \textbf{71}, 125119 (2005)

  \bibitem{HF86} J. E. Hirsch, R. M. Fye, Phys. Rev. Lett
    \textbf{56}, 2521 (1986)

  \bibitem{mem} M. Jarrell, J. E. Gubernatis, Phys. Rep.
    \textbf{269}, 133 (1996)

  \bibitem{fuj84} {{A.}~{Fujimori}}, {{F.}~{Minami}},
    {{S.}~{Sugano}}, {Phys. Rev. B} \textbf{{29}}, {5225} ({1984})

  \bibitem{linio} {{J.}~{van Elp}}, {{H.}~{Eskes}}, {{P.}~{Kuiper}},
    {{G.~A.} {Sawatzky}}, {Phys. Rev. B} \textbf{{45}}, {1612}
    ({1992})

  \bibitem{FSK04} S. V. Faleev, M. van Schilfgaarde, T. Kotani,
    Phys. Rev. Lett.  \textbf{93}, 126406 (2004)

  \bibitem{LR05} J.-L. Li, G.-M. Rignanese, S. G. Louie, Phys.
    Rev. B \textbf{71}, 193102 (2005)

  \bibitem{shar09} S. Sharma, S. Shallcross, J.~K. Dewhurst,
    E.~K.~U. Gross, arXiv:0912.1118

  \bibitem{kob08} S. Kobayashi, Y. Nohara, S. Yamamoto, T.
    Kujiwara, Phys. Rev. B \textbf{70}, 155112 (2008)

  \bibitem{rod09} C. R\"odl, F. Fuchs, J. Furthm\"uller, F.
    Bechstedt, Phys. Rev. B \textbf{79}, 235114 (2009)

  \bibitem{ary95} {{F.}~{Aryasetiawan}}, {{O.}~{Gunnarsson}},
    {Phys. Rev. Lett.} \textbf{{74}}, {3221} ({1995})

  \bibitem{coco05} M. Cococcioni, S. De Gironcoli, Phys. Rev. B
    \textbf{71}, 035105 (2005)

  \bibitem{ren06} {{X.}~{Ren}}, {{I.}~{Leonov}}, {{G.}~{Keller}},
    {{M.}~{Kollar}}, {{I.}~{Nekrasov}}, {{D.}~{Vollhardt}},
    {Phys. Rev. B} \textbf{{74}}, {195114} ({2006})

  \bibitem{man94} {{F.} {Manghi}}, {{C.} {Calandra}},
    {{S.}~{Ossicini}}, {Phys. Rev. Lett.} \textbf{{73}}, {3129}
    ({1994})

  \bibitem{eder07} R. Eder, Phys. Rev. B \textbf{76}, 241103 (2007)

  \bibitem{yin08} Q. Yin, A. Gordienko, X. Wan, S.~Y. Savrasov,
    Phys. Rev. Lett.  \textbf{100}, 066404 (2008)

  \bibitem{miu08} O. Miura, T. Fujiwara, Phys. Rev. B \textbf{77},
    195124 (2008)

  \bibitem{JWA} This conceptual picture was suggested to one of us
    (JK) by J.~W. Allen.

  \bibitem{ishida10} H. Ishida, A. Liebsch, Phys. Rev. B {\bf 81},
    054513 (2010)

  \bibitem{laue09} A. Laeuchli, P. Werner, Phys. Rev. B
    \textbf{80}, 235117 (2009)


  \bibitem{yoo-05} C. S. Yoo, B. Maddox, J.-H. P. Klepeis, V. Iota, W.
    Evans, A. McMahan, M. Y. Hu, P. Chow, M. Somayazulu, D. H\"
    ausermann, R. T. Scalettar, W. E. Pickett, Phys. Rev. Lett.
    \textbf{94}, 115502 (2005)

  \bibitem{GSL08} A. G. Gavriliuk, V. V. Struzhkin, I. S. Lyubutin, S.
    G. Ovchinnikov, M. Y. Hu, P. Chow, Phys. Rev. B \textbf{77},
    155112 (2008)

  \bibitem{lyu09} I. S. Lyubutin, S. G. Ovchinnikov, A. G. Gavriliuk,
    V. V. Struzhkin, Phys. Rev. B \textbf{79}, 085125 (2009)

  \bibitem{cohen97}
    R. E. Cohen, I. I. Mazin, D. G. Isaak, 
    Science {\bf 275}, 654 (1997)

  \bibitem{Sgull-51} C. G. Shull, W. A. Strauser, E. O. Wollan,
    Phys. Rev.  \textbf{83}, 333 (1951)

  \bibitem{Fujimori-86} A. Fujimori, M. Saeki, N. Kimizuka, M.
    Taniguchi, S. Suga, Phys. Rev. B \textbf{34}, 7318 (1986)

  \bibitem{KK02} C.-Y. Kim, C.-Y. Kim, M. J. Bedzyk, E. J. Nelson, J.
    C. Woicik, L. E. Berman, Phys. Rev. B \textbf{66}, 085115
    (2002)

  \bibitem{lad-89} R.~J. Lad, V.~E. Henrich, Phys. Rev. B
    \textbf{39}, 13478 (1989)

  \bibitem{mochizuki} S. Mochizuki, Phys. Status Solidi A \textbf
    {41}, 591 (1977)

  \bibitem{KL85} K.-H. Kim, S.-H. Lee, J.-S. Choi, J. Phys. Chem.
    Solids \textbf{46}, 331 (1985)

  \bibitem{Pasternak-99} M. P. Pasternak, G. Kh. Rozenberg, G. Yu.
    Machavariani, O. Naaman, R. D. Taylor, R. Jeanloz, Phys. Rev.
    Lett. \textbf{82}, 4663 (1999)

  \bibitem{Rozenberg-02} G. Kh. Rozenberg, L. S. Dubrovinsky, M. P.
    Pasternak, O. Naaman, T. Le Bihan, R. Ahuja, Phys. Rev. B
    \textbf{65}, 064112 (2002)

  \bibitem{Liu-03} H. Liu, W. A. Caldwell, L. R. Benedetti, W. Panero,
    R. Jeanloz, Phys. Chem. Miner. \textbf{30}, 582 (2003)

  \bibitem{Badro-02} J. Badro, G. Fiquet, V. V. Struzhkin, M.
    Somayazulu, H.-K. Mao, G. Shen, T. Le Bihan, Phys. Rev. Lett.
    \textbf{89}, 205504 (2002)

  \bibitem{noguchi} Y. Noguchi, K. Kusaba, K. Fukuoka, Y. Syono,
    Geophys. Res. Lett. \textbf{23}, 1469 (1996)

  \bibitem{mita} Y. Mita, Y. Sakai, D. Izaki, M. Kobayashi, S. Endo,
     S. Mochizuki, phys. stat. sol. (b) \textbf{223}, 247 (2001)

  \bibitem{mita2} Y. Mita, D. Izaki, M. Kobayashi, S. Endo, Phys.
    Rev. B \textbf{71}, 100101 (2005)

  \bibitem{patterson} J. R. Patterson, C. M. Aracne, D. D. Jackson, V.
    Malba, S. T. Weir, P. A. Baker, Y. K. Vohra, Phys. Rev. B
    \textbf{69}, 220101 (2004)

  \bibitem{yoo} C.~S. Yoo, B.~R. Maddox, J.-H.~P. Klepeis, V. Iota, W.
    Evans, A. McMahan, M. Hu, P. Chow, M. Somayazulu, D.
    H{\"a}usermann, R.~T. Scalettar, W.~E. Pickett, Phys. Rev.
    Lett. \textbf{94}, 115502 (2005)

  \bibitem{rueff} J.-P. Rueff, A. Mattila, J. Badro, G.
    Vank$\grave{o}$, A. Shukla, J. Phys.: Cond. Matt. \textbf{17},
    S717 (2005)

  \bibitem{werner-07} P. Werner, A.~J. Millis, Phys. Rev. Lett.
    \textbf{99}, 126405 (2007)

  \bibitem{xu01} W. M. Xu, O. Naaman, G.~Kh. Rozenberg, M.~P.
    Pasternak, R.~D. Taylor Phys. Rev. B \textbf{64}, 094411
    (2001)

  \bibitem{ama09} D.~J. Adams, B. Amadon, Phys. Rev. B \textbf{79},
    115114 (2009)

  \bibitem{DM09+KS09} The problem of equilibrium volume of number of
    simple elements has been also recently addressed in
    I. Di Marco,
    J. Min\'{a}r, S. Chadov, M. I. Katsnelson, H. Ebert, A. I.
    Lichtenstein, Phys. Rev. B \textbf{79}, 115111 (2009), and
    A. Kutepov, S. Y. Savrasov, G. Kotliar, Phys. Rev. B
    \textbf{80}, 041103 (2009).

  \bibitem{JT37} H. A. Jahn, E. Teller, Proc. R. Soc. London Ser. A
    \textbf{%
      161}, 220 (1937)

  \bibitem{KK1} D. I. Khomskii, K. I. Kugel, Solid State Comm.
    \textbf{13}, 763 (1973)

  \bibitem{KK2} K. I. Kugel, D. I. Khomskii, Sov. Phys. Solid State
    \textbf{17}, 285 (1975)

  \bibitem{KK3} K. I. Kugel, D. I. Khomskii, Sov. Phys. JETP
    \textbf{52}, 501 (1981)

  \bibitem{KK4} K. I. Kugel, D. I. Khomskii, Sov. Phys. Usp.
    \textbf{25}(4), 231 (1982)

  \bibitem{LNMTO1} O. K. Andersen, Phys. Rev. B \textbf{12}, 3060
    (1975)

  \bibitem{LNMTO2} O. K. Andersen, T. Saha-Dasgupta, Phys. Rev. B
    \textbf{62}, R16219 (2000)

  \bibitem{LB08} I. Leonov, N. Binggeli, Dm. Korotin, V. I. Anisimov,
    N.  Stoji\' c, D. Vollhardt, Phys. Rev. Lett. \textbf{101},
    096405 (2008)

  \bibitem{LK10} I. Leonov, Dm. Korotin, N. Binggeli, V. I. Anisimov,
     D. Vollhardt, Phys. Rev. B \textbf{81}, 075109 (2010)

  \bibitem{TL08} G. Trimarchi, I. Leonov, N. Binggeli, Dm. Korotin,
     V. I. Anisimov, J. Phys.: Condens. Matter \textbf{20}, 135227
    (2008)

  \bibitem{DK08} Dm. Korotin, A. V. Kozhevnikov, S. L. Skornyakov, I.
    Leonov, N. Binggeli, V. I. Anisimov, G. Trimarchi, Eur. Phys.
    J. B \textbf{65}, 91 (2008)

  \bibitem{AL08} B. Amadon, F. Lechermann, A. Georges, F. Jollet, T.
    O. Wehling, A. I. Lichtenstein, Phys. Rev. B \textbf{77},
    205112 (2008)

  \bibitem{LG06} For a formulation of LDA+DMFT within a mixed-basis
    pseudopotential approach see F. Lechermann, A. Georges, A.
    Poteryaev, S. Biermann, M. Posternak, A. Yamasaki, O. K.
    Andersen, Phys. Rev. B \textbf{74}, 125120 (2006)

  \bibitem{entropy} To describe the thermodynamics of solids one must,
    in principle, also compute the entropy and consider the
    electronic, magnetic and lattice (vibrational) contributions in
    the Gibbs free energy. The first two contributions are generally
    small and can be neglected in the paramagnetic phase of a
    wide-band insulator, whereas the lattice entropy may have an
    influence on a structural phase transition.  To estimate its
    contribution would require to perform molecular dynamics
    calculations for a correlated system. This is a very demanding
    project which we plan to do in the future.

  \bibitem{KY67} S. Kadota, I. Yamada, S. Yoneyama, K. Hirakawa,
    J. Phys.  Soc. Jpn. {\bf 23}, 751 (1967)

  \bibitem{BM90} R. H. Buttner, E. N. Maslen, N. Spadaccini, Acta
    Cryst. B \textbf{46}, 131 (1990)

  \bibitem{G63} J. B. Goodenough, \textit{Magnetism and the Chemical
      Bond} (Interscience, New York, 1963)

  \bibitem{MK02} J. E. Medvedeva, M. A. Korotin, V. I. Anisimov,
    A. J. Freeman, Phys. Rev. B \textbf{65}, 172413 (2002)

  \bibitem{PK08} E. Pavarini, E. Koch, A. I. Lichtenstein, Phys.
    Rev.  Lett. \textbf{101}, 266405 (2008)

  \bibitem{BA04} N. Binggeli, M. Altarelli, Phys. Rev. B
    \textbf{70}, 085117 (2004)

  \bibitem{GGALDA} In general, GGA tends to give better results than
    LDA for the electronic and structural properties of complex oxides
    and related materials. See, D. R. Hamann, Phys. Rev. Lett.
    \textbf{76}, 660 (1996) and H. Sawada, Y. Morikawa, K. Terakura,
     N. Hamada, Phys. Rev. B \textbf{56}, 12154 (1997)

  \bibitem{HS69} M. T. Hutchings, E. J. Samuelsen, G. Shirane, K.
    Hirakawa, Phys. Rev. \textbf{188}, 919 (1969)

  \bibitem{U91} T. Ueda, K. Sugawara, T. Kondo, I. Yamada, Solid
    State Commun. \textbf{80}, 801 (1991)

  \bibitem{Y89} I. Yamada, H. Fujii, M. Hidaka, J. Phys. Condens.
    Matter \textbf{1}, 3397 (1989)

  \bibitem{EZ08} M. V. Eremin, D. V. Zakharov, H.-A. Krug von Nidda,
    R. M.  Eremina, A. Shuvaev, A. Pimenov, P. Ghigna, J. Deisenhofer,
     A. Loidl, Phys. Rev. Lett. \textbf{101}, 147601 (2008)

  \bibitem{DL08} J. Deisenhofer, I. Leonov, M.  V. Eremin, Ch. Kant,
    P. Ghigna, F. Mayr, V. V. Iglamov, V. I. Anisimov, D. van der
    Marel, Phys. Rev. Lett. \textbf{101}, 157406 (2008)

  \bibitem{PC02} L. Paolasini, R. Caciuffo, A. Sollier, P. Ghigna,
    M.  Altarelli, Phys. Rev. Lett. \textbf{88}, 106403 (2002)

  \bibitem{CP02} R. Caciuffo, L.  Paolasini, A. Sollier, P. Ghigna, E.
    Pavarini, J. van den Brink, M.  Altarelli, Phys. Rev. B
    \textbf{65}, 174425 (2002)

  \bibitem{PSEUDO} Calculations have been performed using the Quantum
    ESPRESSO package, see Ref.~\onlinecite{RevPWSCF1,RevPWSCF2},
    \url{http://www.quantum-espresso.org}

  \bibitem{dif_orb} The local coordinate system is chosen with the
    \emph{z} direction defined along the longest (in $ab$ plane) Cu-F
    bond of the CuF$_6$ octahedron.

  \bibitem{full_selfconsistency} Here we perform DMFT calculations for
    a fixed DFT Hamiltonian $\hat H_{DFT}$, thereby neglecting full
    charge self-consistency which is not expected to change the
    results significantly~\cite{AB06}.

  \bibitem{Off-diag-elements} To simplify the computation we neglected
    the orbital off-diagonal elements of the local Green function by
    applying an additional transformation into the local basis set
    with a diagonal density matrix during each DMFT iteration.

  \bibitem{EL71} J. B. A. A. Elemans, B. van Laar, K. R. van der
    Veen, B. O. Loopstra, J. Phys. Chem. Solids \textbf{3}, 238
    (1971)

  \bibitem{RH98} J. Rodriguez-Carvajal, M. Hennion, F. Moussa, A. H.
    Moudden, L. Pinsard, A. Revcolevschi, Phys. Rev. B
    \textbf{57}, R3189 (1998)

  \bibitem{CF03} T. Chatterji, F. Fauth, B. Ouladdiaf, P. Mandal,
    B.  Ghosh, Phys. Rev. B \textbf{68}, 052406 (2003)

  \bibitem{TB05} G. Trimarchi, N. Binggeli, Phys. Rev. B
    \textbf{71}, 035101 (2005)

  \bibitem{PZ00} Th. Pruschke, M. B. Z\"olfl, \emph{Advances in
      Solid State Physics} \textbf{40}, 251 (2000); See also R. Peters,
    Th. Pruschke, cond-mat/0908.3990, where the authors discuss
    the interplay of orbital and spin degrees of freedom in the two
    orbital Hubbard model near quarter filling.

  \bibitem{YF06} A. Yamasaki, M. Feldbacher, Y.-F. Yang, O. K.
    Andersen, K. Held, Phys. Rev. Lett. \textbf{96}, 166401 (2006)

  \bibitem{HA08} K. Held, O. K.  Andersen, M. Feldbacher, A. Yamasaki,
     Y.-F. Yang, J. Phys.: Condens.  Matter \textbf{20}, 064202
    (2008)

  \bibitem{PK09} E. Pavarini, E. Koch, cond-mat/0904.4603

  \bibitem{YV06} W.-G. Yin, D. Volja, W. Ku, Phys. Rev. Lett.
    \textbf{96}, 116405 (2006)

  \bibitem{dif_orb_lamno} The local coordinate system is chosen such
    that the GGA Mn $3d$ density matrix has a diagonal form.

  \bibitem{MS96} A. J. Millis, B. I. Shraiman, R. Mueller, Phys.
    Rev.  Lett. \textbf{77}, 175 (1996)

  \bibitem{HV00} K. Held, D. Vollhardt, Phys. Rev. Lett.
    \textbf{84}, 5168 (2000)

  \bibitem{E6:Nekrasov2006a} I.~A. Nekrasov, K.~Held, G.~Keller, D.~E.
    Kondakov, T.~Pruschke, M.~Kollar, O.~K. Andersen, V.~I. Anisimov,
     D.~Vollhardt, Phys. Rev. B \textbf{73}, 155112 (2006)

  \bibitem{Gunnarsson89} O. Gunnarsson, O. K. Andersen, O. Jepsen,
    J. Zaanen, Phys. Rev. B \textbf{39}, 1708 (1989)
  
  \bibitem{E6:Byczuk07a} K.~Byczuk, M.~Kollar, K.~Held, Y.-F. Yang,
    I.~A. Nekrasov, T.~Pruschke, D.~Vollhardt, Nature Physics
    \textbf{3}, 168 (2007)

    \bibitem{Yoshida05} T. Yoshida, K. Tanaka, H. Yagi, A. Ino, H.
    Eisaki, A. Fujimori, Z.-X. Shen, Phys. Rev. Lett. {\bf 95}, 146404
    (2005)

  \bibitem{Takizawa09} M. Takizawa, M. Minohara, H. Kumigashira, D.
    Toyota, M. Oshima, H. Wadati, T. Yoshida, A. Fujimori, M. Lippmaa,
    M. Kawasaki, H. Koinuma, G. Sordi, M. Rozenberg, Phys. Rev. B {\bf
      80}, 235104 (2009)
 
\bibitem{Toschi2009}
A. Toschi, M. Capone, C. Castellani, K. Held, Phys. Rev. Lett. {\bf 102},
  076402 (2009)


  \bibitem{E6:Uhrig2009}
  C. Raas, P. Grete, G. S.  Uhrig, Phys. Rev. Lett. {\bf 102},
  076406 (2009)


 \bibitem{E6:Nickel2009}
  A. Hofmann, X. Y. Cui, J. Sch\"{a}fer, S. Meyer, P. H\"{o}pfner, C.
  Blumenstein, M. Paul, L. Patthey, E. Rotenberg, J. B\"{u}nemann, F.
  Gebhard, T. Ohm, W. Weber, R. Claessen, Phys. Rev. Lett.
  {\bf 102}, 187204 (2009)

  \end{thebibliography}
\end{document}